\documentclass[reprint,amsmath,amssymb,aip]{revtex4-1}
\usepackage{graphicx}
\usepackage{dcolumn}
\usepackage{bm}
\usepackage[usenames,dvipsnames]{color}

\def \be {\begin{equation}}
\def \ee {\end{equation}}
\def \ben {\begin{eqnarray}}
\def \een {\end{eqnarray}}
\def \bi {\begin{itemize}}
\def \ei {\end{itemize}}
\def\re#1{{\color{black} {#1}}}
\def\ret#1{{\color{black} {#1}}}
\def\rc#1{{\color{black} {#1}}}
\usepackage{multirow}
\usepackage{ctable}

\begin{document}
\title{Fermi's golden rule rate expression for transitions due to nonadiabatic derivative couplings in the adiabatic basis}

\author{Seogjoo J. Jang}
\email{seogjoo.jang@qc.cuny.edu}

\affiliation{Department of Chemistry and Biochemistry, Queens College, City University of New York, 65-30 Kissena Boulevard, Queens, New York 11367, USA \& PhD Programs in Chemistry and Physics, Graduate Center of the City University of New York, New York 10016, USA }

\author{Byeong Ki Min}
\affiliation{Department of Chemistry, Korea Advanced Institute of Science and Technology, Daejeon 34141, Korea}

\author{Young Min Rhee}
\affiliation{Department of Chemistry, Korea Advanced Institute of Science and Technology, Daejeon 34141, Korea}

\date{Published in the {\it Journal of Chemical Theory and Computation} {\bf 21}, 1850-1864 (2025) }

\begin{abstract}
Starting from a general molecular Hamiltonian expressed in the basis of adiabatic electronic and nuclear position states, where a compact and complete expression for the nonadiabatic derivative coupling (NDC) Hamiltonian term is obtained, we provide a general analysis of the Fermi’s golden rule (FGR) rate expression for nonadiabatic transitions between adiabatic states. We then consider a quasi-adiabatic approximation that uses crude adiabatic states and NDC couplings, both evaluated at the minimum potential energy configuration of the initial adiabatic state, for the definition of the zeroth and first-order terms of the Hamiltonian. Although the application of this approximation is rather limited, it allows deriving a general FGR rate expression without further approximation while accounting for non-Condon contribution to the FGR rate arising from momentum operators of NDC terms and its coupling with vibronic displacements. For a generic and widely used model where all nuclear degrees of freedom and environmental effects are represented as linearly coupled harmonic oscillators, we derive a closed-form FGR rate expression that requires only Fourier transform. The resulting rate expression includes quadratic contributions of NDC terms and their couplings to Franck–Condon modes, which require evaluation of two additional bath spectral densities in addition to the conventional one that appears in a typical FGR rate theory based on the Condon approximation. Model calculations for the case where nuclear vibrations consist of both a sharp high-frequency mode and an Ohmic bath spectral density illustrate new features and implications of the rate expression. We then apply our theoretical expression to the nonradiative decay from the first excited singlet state of azulene, which illustrates the utility and implications of our theoretical results.
\end{abstract}

\maketitle

\section{Introduction}
Advances in electronic structure calculation and quantum dynamics methods over past decades have made it possible to conduct first principles dynamics calculation for many molecular systems.\cite{makhov-cp493,tao-jpca113,goings-wcms2017,wang-arpc66,curchod-cr118,guo-jcp155,zhao-jpcl11,song-jctc16,hegger-jctc16,gao-acr55}  As yet, there remain significant challenges for accurate quantum dynamics calculations of  excited electronic states, \ret{especially for molecules in condensed or complex environments}.   One crucial issue that has to be addressed carefully in this regard is the fact that most quantum dynamics methods and rate theories have been developed under the assumption that it is possible to identify diabatic electronic states with constant or simple forms of electronic couplings between them.  On the other hand, most quantum chemistry methods first seek for the calculation of adiabatic electronic states for fixed nuclei, although alternative approaches are being developed and extended.\cite{deumens-rmp66,abedi-prl105,abedi-jcp137,min-jpcl8,he-wires12,wu-jpcl15-1,he-jpcl15}  In addition, \rc{accurately} incorporating the effects of environmental dynamics into otherwise very high quality {\it ab initio} calculations remains challenging. 

In general, there is no genuine unitary transformation (independent of nuclear coordinates) from an arbitrary adiabatic basis to a true diabatic one.  Thus, it remains an important practical issue to develop dynamics methods and rate theories starting from adiabatic states as little assumptions as possible.   To this end, \rc{full} characterization of nonadiabatic derivative coupling (NDC) terms between adiabatic states is needed.  In fact, theoretical works on this issue have a long history.  \rc{Important formal developments} have already been made.\cite{kubo-pr86,kubo-ptp13,lin-jcp44,lin-jcp58,nitzan-jcp56,fujimura-jcp66,mebel-jpca103,niu-jpca114,peng-cp370,wang-jcp154,borrelli-jcp129,sharf-cpl9,orlandi-cpl8}  However, considering recent computational and experimental advances, it is meaningful to reassess \rc{ issues addressed in earlier theoretical works }in the context of modern computational modeling of  nonadiabatic transitions.  

While consideration in an adiabatic basis \rc{is straightforward} for conducting nuclear quantum dynamics \re{on a single adiabatic electronic surface}, extension of such approach for electron-nuclear dynamics involving multiple adiabatic states, \ret{especially in condensed or complex environments}, is challenging. \ret{To this end,} various approximate methods\cite{makhov-cp493,tao-jpca113,goings-wcms2017,wang-arpc66,curchod-cr118,guo-jcp155,zhao-jpcl11,song-jctc16,tully-jcp137,jasper-acr39,izmaylov-jcp135,kapral-arpc57,subotnik-arpc67,esch-jcp155,huang-jcp159,zhu-jcp137,joubert-doriol-jpca122,zhou-jpcl10,prezhdo-acr54,shu-jctc18,min-jpcl8,wu-jpcl15-1,he-jpcl15} have been developed.  As yet, there \re{are} two important theoretical issues that need more careful theoretical consideration even for cases where the dynamics can be modeled as rate processes.  One is the nonorthogonality of different adiabatic electronic states for different values of nuclear coordinates and the other is complicated nature of couplings between them.   These issues are also important  for \rc{proper} modeling of excitons formed in groups of molecules.  Historically, Frenkel-type exciton-bath models\cite{knox,kenkre-reineker,may} \rc{have been used with reasonable success} for describing many experimental data.  However, the diabatic states used to define local site excitation states in these models are not always clearly defined.\cite{jang-exciton}  In addition, the extent of how and in what ways NDC terms contribute to the properties of excitons  for many systems remain open issues.   
 
In this work, we carefully consider NDC terms between adiabatic electronic states and  provide a general FGR rate expression for nonadiabatic transitions between adiabatic states, which incorporates expressions used in many of earlier theories\cite{kubo-pr86,kubo-ptp13,lin-jcp44,lin-jcp58,mebel-jpca103,niu-scc51,niu-jpca114,wang-jcp154,borrelli-jcp129} in a compact manner.  \ret{Much of this amounts to a reformulation of already known theories but offers a new perspective.}  We then consider a well-defined Fermi's golden rule (FGR) rate expression under a quasi-adiabatic approximation, and provide a \ret{new closed form FGR rate expression.  This expression clearly accounts for non-Condon effects due to momentum contribution to the FGR rate expression and \rc{can be evaluated employing the data available from the }standard {\it ab initio} and dynamics calculation for vibrational/environmental relaxation dynamics.}

\section{Fermi's golden rule rate expression for nonadiabatic transitions between adiabatic states}
Let us first provide a brief overview of adiabatic states and nonadiabatic couplings.  Although this is a standard topic of quantum chemistry, notations and definitions provided in this section are based on a more complete consideration\cite{jang-jcp137} and are different from conventional ones. In addition, the overview here will help clarify issues implicit in applying the standard FGR rate expression for nonadiabatic transitions and will also provide a compact formalism for our rate expression.  For a complete exposition including textbook level explanations, readers can refer to the Supporting Information (SI).
   
\re{Consider a molecular system consisting of $N_u$ nuclei, with their positions collectively represented by a $3N_u$ dimensional vector ${\bf R}$, and assume that it is possible to identify two major adiabatic electronic states,
\cite{Note1} $|\psi_{e,1}({\bf R})\rangle$ and $|\psi_{e,2} ({\bf R} )\rangle$, which are well separated from other adiabatic electronic states.  We assume that these two states have the same spin multiplicity and thus do not consider spin states explicitly here.  Then, as detailed in SI, the total molecular Hamiltonian in this subspace can be expressed as }
\be
\hat H=\hat H_{ad,1}+\hat H_{ad,2}+\frac{1}{2}\sum_{\alpha=1}^{3N_u} \left (\hat P_\alpha\hat F_\alpha+\hat F_\alpha \hat P_\alpha \right)+\hat S ,  \label{eq:hamil-mol-2}
\ee
where $\hat H_{ad,1}$ and $\hat H_{ad,2}$ are adiabatic components of the Hamiltonian, $\hat P_\alpha$ is the (one dimensional) nuclear momentum operator along the $\alpha$ direction, $\hat F_\alpha$ is the operator representing the first nonadiabatic derivative coupling (NDC) terms, and 
$\hat S$ represents the second NDC terms.  \re{Note that the only approximation involved in eq. \ref{eq:hamil-mol-2} is the finite truncation of the electronic Hilbert space.  }   

In more detail, the first two terms of eq. \ref{eq:hamil-mol-2}, for $k=1$, $2$,  are 
\ben
&&\hat H_{ad,k}=\int d{\bf R} |{\bf R}\rangle |\psi_{e,k} ({\bf R})\rangle \Big \{ -\sum_{\alpha=1}^{3N_u} \frac{\hbar^2}{2M_\alpha} \frac{\partial^2}{\partial R_\alpha^2}  \nonumber \\ &&\hspace{.2in}+U_k({\bf R})\Big \}\langle \psi_{e,k} ({\bf R})|\langle {\bf R}| ,
\een
where $|{\bf R}\rangle$ is the nuclear position state, $M_\alpha$ is the nuclear mass associated with the momentum along the $\alpha$ direction, $U_k({\bf R})$ is the sum of  the eigenvalue $E_k({\bf R})$ of the electronic state $|\psi_{e,k}({\bf R})\rangle$ plus the nuclear potential energy terms.  See SI for a more detailed expression for $U_k({\bf R})$.    Note that $|{\bf R}\rangle|\psi_{e,k}({\bf R})\rangle$ (or $\langle \psi_{e,k}({\bf R})|\langle {\bf R}|$)  is a short notation for a direct product of electronic and nuclear states.  This simple direct product form involving adiabatic electronic state is possible only for nuclear position state for which an adiabatic electronic state is unambiguously defined.

As described (for more general case) in SI, the first NDC terms in eq. \ref{eq:hamil-mol-2} are expressed as
\be
\hat F_\alpha=\int d{\bf R} \sum_{k=1}^2\sum_{k'=1}^2 |{\bf R}\rangle |\psi_{e,k}({\bf R})\rangle F_{\alpha,kk'} ({\bf R}) \langle \psi_{e,k'} ({\bf R})|\langle {\bf R}|  , \label{eq:falpha}
\ee
with
\be
F_{\alpha,kk'} ({\bf R}) =\frac{\hbar}{iM_\alpha} \langle \psi_{e,k} ({\bf R})| \left ( \frac{\partial}{\partial R_\alpha} |\psi_{e,k'}({\bf R})\rangle \right)  , \label{eq:falpha-kkp}
\ee
and the second NDC term in eq. \ref{eq:hamil-mol-2} is expressed as
\be
\hat S =\int d{\bf R} \sum_{k=1}^2\sum_{k'=1}^2 |{\bf R}\rangle |\psi_{e,k} ({\bf R})\rangle S_{kk'} ({\bf R}) \langle \psi_{e,k'} ({\bf R}) |\langle {\bf R}|,  
\ee
with 
\be
S_{kk'}({\bf R})=\frac{1}{2}\sum_{\alpha=1}^{3N_u} \sum_{k''=1}^2  F_{\alpha,kk''} ({\bf R} ) F_{\alpha,k''k'}({\bf R}) . \label{eq:skkp}
\ee
\re{The definitions given by eqs. \ref{eq:falpha-kkp} and \ref{eq:skkp} make it clear that $\hat H$ defined by eq. \ref{eq:hamil-mol-2} is indeed Hermitian.  It is also important to note that $S_{kk'}({\bf R})$ contains only products of the first derivatives of adiabatic electronic eigenstates with respect to nuclear coordinates since second derivatives are already contained in $\hat P_\alpha\hat F_\alpha$ of eq. \ref{eq:hamil-mol-2}. }

For further simplification, let us assume that real valued eigenfunctions for $\langle {\bf r}|\psi_{e,k}({\bf R})\rangle$ can be identified, where ${\bf r}$ represents collectively positions of all the electrons and $\langle {\bf r}|$ the corresponding position bra.   This \ret{assumption is valid for transitions involving }singlet states in the absence of magnetic field.  Under this assumption,  $F_{\alpha,11}({\bf R})=F_{\alpha,22}({\bf R})=0$ and  $S_{12}({\bf R})=S_{21}({\bf R})=0$.   For these cases, the second NDC terms are diagonal in the adiabatic basis and can be combined into the adiabatic terms of the Hamiltonian  to define a zeroth order Hamiltonian as follows: 
\be
\hat H_0=\hat H_{0,1}+\hat H_{0,2} , \label{eq:h0}
\ee
where $\hat H_{0,k}=\hat H_{ad,k} +\hat S_k$ with
\be
\hat S_{k}=\int d{\bf R} |{\bf R}\rangle |\psi_{e,k} ({\bf R})\rangle S_{kk}({\bf R}) \langle \psi_{e,k} ({\bf R})|\langle {\bf R}| .
\ee
Then, the full molecular Hamiltonian, eq. \ref{eq:hamil-mol-2}, can also be expressed as
\be
\hat H=\hat H_0+\hat H_c ,\label{eq:hamil-2}
\ee
where $\hat H_c$ contains only the first NDC terms.  This is offdiagonal with respect to adiabatic states and is given by
\be
\hat H_c=\frac{1}{2}\sum_{\alpha=1}^{3N_u} \left (\hat P_\alpha\hat F_\alpha+\hat F_\alpha\hat P_\alpha \right)  . \label{eq:hc-0}
\ee

Having defined the zeroth order Hamiltonian $\hat H_0$ and  $\hat H_c$, eqs. \ref{eq:h0} and \ref{eq:hc-0}, it now \rc{seems} straightforward to employ the following standard FGR rate expression:
\be
k_{_{FGR}}=\frac{2\pi}{\hbar}\sum_{i,f} p_i\left |\langle \psi_f|\hat H_c|\psi_i\rangle \right|^2 \delta (E_f-E_i)  ,\label{eq:k-fgr-std-1}
\ee 
where $|\psi_i\rangle$ and  $|\psi_f\rangle$ are initial and final eigenstates of the zeroth order Hamiltonian $\hat H_0$ with eigenvalues $E_i$ and $E_f$, and $p_i$ is the probability for the initial state $|\psi_i\rangle$.   \rc{However, exact application of eq. \ref{eq:k-fgr-std-1} still involves somewhat complicated theoretical issues as detailed below. }  

\rc{Most theories\cite{kubo-pr86,kubo-ptp13,lin-jcp44,lin-jcp58,mebel-jpca103,fujimura-jcp66,niu-jpca114,peng-cp370}  of nonadiabatic transition have considered eq. \ref{eq:k-fgr-std-1} as the starting point.  However, with few exceptions,\cite{nitzan-jcp56,sharf-cpl9,orlandi-cpl8} they used} approximate expressions for the coupling obtained from nonadiabatic corrections of adiabatic vibronic wavefunctions.   On the other hand, \rc{for proper application of eq. \ref{eq:k-fgr-std-1},} it is also important to identify the initial and final states that are genuine orthogonal eigenstates of $\hat H_0$, \rc{while being} {\it defined  in the full direct product space of electronic and nuclear degrees of freedom}.  The issue at hand becomes clearer considering the following time domain expression for FGR:   
\be
k_{_{FGR}}=\frac{1}{\hbar^2}\sum_f \int_{-\infty}^\infty dt \langle \psi_f |e^{i\hat H_0 t/\hbar} \hat H_c e^{-i\hat H_0 t/\hbar}\hat \rho_i \hat H_c |\psi_f\rangle , \label{eq:k-fgr-std}
\ee
where  $\hat \rho_i=\sum_i p_i|\psi_i\rangle\langle \psi_i|$. The above expression results directly from the  first order expansion of the propagator with respect to $\hat H_c$ as follows: 
 \be
 e^{-i\hat Ht/\hbar}\approx \int_0^t d\tau e^{-it\hat H_0 (t-\tau)/\hbar}\hat H_c  e^{-it\hat H_0 \tau/\hbar} ,  \label{eq:1st-order-app}
 \ee
 and \rc{results in} eq. \ref{eq:k-fgr-std-1} only if $\hat H_0|\psi_i\rangle=E_i|\psi_i\rangle$ and $\hat H_0|\psi_f\rangle=E_f|\psi_f\rangle$.   \rc{If not, eq. \ref{eq:k-fgr-std-1} amounts to invoking an additional approximation.}
 
 In SI, we prove directly that the following state indeed is an eigenstate of $\hat H_{ad,k}$:
 \ben
|\Psi_{k,n}\rangle =\int d{\bf R} \     \chi_{n_k}({\bf R}) |{\bf R}\rangle |\psi_{e,k}( {\bf R})\rangle , k=1,2 , \label{eq:adia-eigen}
\een
where $\chi_{n_k}({\bf R})$ is a nuclear eigenfunction for the adiabatic potential energy, $U_k({\bf R})$. Note that the projection of the above state onto a particular value of nuclear coordinate ${\bf R}'$ results in $\langle {\bf R}'|\Psi_{k,n}\rangle =\chi_{n_k}({\bf R}') |\psi_{e,k}( {\bf R}')\rangle$, which is the conventional adiabatic electronic-nuclear wavefunction.  As yet, $|\Psi_{k,n}\rangle$ is not the eigenstate of $\hat H_0$ unless $\hat S$ is constant, although this effect can be accounted for up  to the first order by using an average value. 

In SI, we also provide a full expression for the matrix elements of $\hat H_c$ with respect to states given by eq. \ref{eq:adia-eigen}, which includes additional terms that have not been considered in earlier theories.   These additional terms \rc{appear because} eq. \ref{eq:adia-eigen} involves a linear combination of adiabatic electronic states that are not orthogonal to each other.  Namely,  \rc{it results from the fact that }$\langle \psi_k({\bf R})|\psi_{k'} ({\bf R}')\rangle\neq 0$ for ${\bf R}\neq {\bf R}'$ even for $k\neq k'$, as long as they have the same spin symmetry. The expressions provided in the SI, \rc{eqs. S20 and S21}, also show that the effect of nuclear coordinate dependence of adiabatic states should be handled \rc{carefully including the second NDC terms as well}. 
 
 The analysis provided above and the full expression for matrix elements of $\hat H_c$ \rc{in SI} clarify \rc{issues concerning  exact calculation of the FGR rate for }nonadiabatic transitions between adiabatic states.   Errors due to making approximations in this calculation can be significant if adiabatic states change significantly with nuclear coordinates and/or $\hat H_c$ is not small enough.  Theories addressing these issues have indeed been developed, but seem to have considered only parts of the matrix elements of $\hat H_c$.\cite{nitzan-jcp56,fujimura-jcp66}   Alternatively, one can employ crude adiabatic states as the reference electronic states, which does not introduce complications resulting from the peculiar nature of  adiabatic states.  Indeed, advanced formulations in this direction, known as crude adiabatic schemes, have already been \rc{formulated}.\cite{sharf-cpl9,orlandi-cpl8} However, these have not yet been developed into practical computational methods to the best of our knowledge.     

Our work here is focused on the case where $\hat H_c$ remains small and changes modestly with nuclear coordinates.   Thus, we here consider a quasiadiabatic approximation at the level of the Hamiltonian operator, which replaces  adiabatic electronic states \rc{involved in} $\hat H_0$ and  $\hat H_c$, eqs \ref{eq:h0} and \ref{eq:hc-0}, with appropriate crude adiabatic states.   This approximation accounts for the nonadiabatic coupling at the lowest order of the Hamiltonian, and makes the application of FGR straightforward and well defined.  This also opens up future possibility to make further theoretical advances following the idea of the crude adiabatic scheme\cite{sharf-cpl9,orlandi-cpl8} and/or determining the correction terms of the Hamiltonian employing perturbative expansion\cite{nitzan-jcp56} of adiabatic states with respect to crude adiabatic states.   

 It is also interesting to note that our general FGR rate expression based on the simple quasiadiabatic approximation turns out to be \ret{almost equivalent (save for the diagonal second NDC term) to the general expression used by Kubo and Toyozawa\cite{kubo-ptp13} and \rc{that}\cite{niu-jpca114,peng-cp370}  used later by Shuai and coworkers (without assumption of promoting mode),} which were derived based on a Condon approximation for the first NDC terms.  A similar expression was used for the calculation of nonradiative decay within the cumulant approximation \ret{by Borrelli and Peluso\cite{borrelli-jcp129} and more recently by Landi {\it et al.}.\cite{landi-jcp160}}  \ret{As will become clear, what is new in our work is that we provide a convenient new form of the FGR rate expression applicable to processes in general environments, which can be calculated by incorporating standard {\it ab initio} data for excited states and the dynamics data for molecular/environmental relaxation}.

\section{Fermi's golden rule rate expression within a quasi-adiabatic approximation}
We here provide the FGR rate expression within the quasi-adiabatic approximation that uses the crude adiabatic states  evaluated at the local minimum of ${\bf R}$ for the initial adiabatic Hamiltonian, $\hat H_{ad,1}$.  More detailed description is provided below.   We also \rc{regard} all other degrees of freedom except for the two electronic states $1$ and $2$ as bath.  \rc{Thus, the bath includes all the molecular vibrations for isolated molecules and also additional environmental effects for molecules embedded in other media.}  

\subsubsection{General expression}
First, consider the case of an isolated molecule. Let us denote the minimum energy nuclear coordinates of the $E_{e,1}({\bf R}$) collectively as ${\bf R}_1^g$.   Then, we define the two adiabatic electronic states determined at ${\bf R}_1^g$  as two electronic states independent of nuclear degrees of freedom as follows: 
\be 
|\psi_{e,k}\rangle =|\psi_{e,k}({\bf R}_1^g)\rangle,\  k=1,2 . 
\ee
It is assumed that these states serve as good approximations for adiabatic electronic states $|\psi_{e,k}({\bf R})\rangle$ near the vicinity of ${\bf R}_1^g$, where nonadiabatic transitions occur.     Thus, the diagonal terms of the Hamiltonian in eq. \ref{eq:hamil-2}  are approximated as\cite{jang-jcp137}
\ben
\hat H_0&\approx&  \Big \{E_1^0+\hat H_{b,1}\Big \}|\psi_{e,1} \rangle\langle \psi_{e,1}|   \nonumber \\
&+&  \Big \{ E_2^0+\hat H_{b,2}\Big \}|\psi_{e,2}\rangle \langle \psi_{e,2} | .\label{eq:h-ad-1}
\een
For isolated molecules, $E_k^0=U_k({\bf R}_1^g) +S_{kk}({\bf R}_1^g)$ and 
\be 
\hat H_{b,k}= \sum_{\alpha=1}^{3N_u} \frac{\hat P_\alpha^2}{2M_\alpha}   +U_k(\hat {\bf R})-U_k({\bf R}_1^g) .
\ee
Full expressions for $E_k^0+\hat H_{b,k}$, both in terms of original nuclear coordinates and normal vibrational modes,  are provided in SI. 

Similarly, it is assumed that the first NDC terms at ${\bf R}_1^g$ also serve as a good approximation for those terms evaluated at nearby nuclear coordinates. 
Thus, the coupling term in eq. \ref{eq:hamil-2} is approximated as 
\ben
\hat H_c&\approx & \sum_{\alpha=1}^{3N_u} \hat P_\alpha \left (F_{\alpha,12}({\bf R}_{1}^g) |\psi_{e,1}\rangle \langle \psi_{e,2}|  \right .  \nonumber \\
&&\left . \hspace{.5in} + F_{\alpha,12}^*({\bf R}_1^g)|\psi_{e,2}\rangle \langle \psi_{e,1} |   \right ) , \label{eq:hc-1}
\een
where $F_{\alpha,12}({\bf R}_{1}^g)$ is defined by eq. \ref{eq:falpha-kkp} evaluated at ${\bf R}_{1}^g$ for $k=1$ and $k'=2$.  In the above equation, the fact that $F_{\alpha,21}({\bf R}_{1}^g)$ is the complex conjugate of $F_{\alpha,12}({\bf R}_{1}^g)$ has also been used.  Note also that $\hat P_\alpha$ \rc{commutes with the diabatic electronic states $|\psi_{e,1}\rangle$ and $|\psi_{e,2}\rangle$}.

 \rc{Note that $S_{11}({\bf R}_g)=S_{22}({\bf R}_g)$ for the present case that has only two adiabatic states. Therefore, these second order NDC terms are assumed not to affect the energy gap. In fact,  the zeroth order Hamiltonian, eq. (\ref{eq:h-ad-1}),} can  include the full second NDC term by replacing   $U_k(\hat {\bf R})$ with $U_k(\hat {\bf R})+S_{kk}(\hat {\bf R})-S_{kk}({\bf R}_1^g)$.  Likewise, $F_{\alpha,12}$ in eq. \ref{eq:hc-1} and its Hermitian conjugate can be assumed to be functions of  nuclear position operators such that each term in the summation is replaced with $\hat P_\alpha \hat F_{\alpha,12}(\hat {\bf R})|\psi_{e,1}\rangle \langle \psi_{e,2}|+ \hat F_{\alpha,21} (\hat {\bf R})\hat P_\alpha |\psi_{e,2}\rangle\langle \psi_{e,1}|$. However, whether this \rc{more general assumption, without fully accounting for the nuclear coordinate dependence of adiabatic states, would consistently} lead to an improvement is not clearly understood yet.

For the Hamiltonian terms $\hat H_0$ and $\hat H_c$ given by eqs. \ref{eq:h-ad-1} and \ref{eq:hc-1} respectively and for the following initial density operator:
\be
\hat \rho_1=|\psi_{e,1}\rangle\langle \psi_{e,1}| \hat \rho_{b,1},
\ee
with $\hat \rho_{b,1}=e^{-\beta \hat H_{b,1}}/Tr\{e^{-\beta\hat H_{b,1}}\}$, it is now straightforward to apply FGR\cite{fermi,jang-jcp159} and obtain an expression similar to well-known expressions.   To this end, it is convenient to use mass weighted coordinate system in the Eckart frame followed by normal mode representation.  Detailed mathematical procedure for this is reviewed in SI.

So far, we  have consider isolated molecules only, but all the expressions can be extended to cases where molecules are embedded in some host environments, which are also detailed in SI. The major change in this case is the extension of the bath Hamiltonians and the normal vibrational modes.  Even in this case, the formal expression for $\hat H_c$ in terms of normal modes remains the same and is given by 
\be
\hat H_c= \sum_{j=1}^{N_{v,1}} \hat p_{1,j} \Big \{ \tilde {F}_{j,12} |\psi_{e,1}\rangle\langle \psi_{e,2}|+\tilde F_{j,12}^* |\psi_{e,2}\rangle\langle \psi_{e,1}| \Big \}  ,  \label{eq:hc-2}
\ee
\re{with
\be
\tilde F_{j,12}=i \hbar\frac{\langle \psi_{e,1}|\left (\partial \hat H_e({\bf R})/\partial  q_{1,j} |_{{\bf R}={\bf R}_1^g} \right )  |\psi_{e,2}\rangle}{U_{1}({\bf R}_1^g)-U_{2}({\bf R}_1^g)} , \label{eq:tf-j}
\ee 
where we have used the fact that $E_1 ({\bf R}_1^g) - E_2 ({\bf R}_1^g) =U_{1}({\bf R}_1^g)-U_{2}({\bf R}_1^g)$. }

In eq. \ref{eq:hc-2}, $\hat p_{1,j}$ is the momentum operator for the $j$th normal mode defined with respect to the minimum potential energy structure of the adiabatic electronic state 1.  As detailed in SI, the above expression, for the case of isolated molecules, is obtained under the assumption that $\hat H_c$ is independent of translation or rotation in the body fixed frame corresponding to the minimum energy structure for the adiabatic electronic state $1$.  For molecules embedded in other media,  this can be justified in more straightforward manner as explained in SI.

\re{Given all the definitions and assumptions, it is now straightforward to apply the time dependent perturbation theory and obtain the following standard FGR rate expression:}\cite{jang-jcp159} 
\be
k_{FGR}=\frac{2}{\hbar^2}{\rm Re} \sum_{j,j'}  \tilde F_{j,12}^*\tilde F_{j',12} \int_0^\infty dt e^{i(E_2^0-E_1^0)t/\hbar} {\mathcal C}_{jj'}(t) , \label{eq:fgr-1p}
\ee
where 
\be
{\mathcal C}_{jj'}(t)={\rm Tr}_b \left \{e^{i t\hat H_{b,2}/\hbar} \hat p_{1,j} e^{-i t\hat H_{b,1}/\hbar} \hat \rho_{b,1} \hat p_{1,j'} \right\} , \label{eq:cjj't}
\ee
\re{with $Tr_b$ representing trace over all the bath degrees of freedom. } 

Equation \ref{eq:fgr-1p} with eq. \ref{eq:cjj't} is the most general expression based on the quasi-adiabatic approximation. It is interesting to note that this expression is equivalent to the general expression derived by Shuai and coworkers\cite{niu-jpca114,peng-cp370} based on a Condon approximation for the first NDC term.   \ret{In fact, a similar expression has been presented much earlier by Kubo and Toyozawa in their generating function formalism for radiative and nonradiative process involving trapped electron in crystal environments.\cite{kubo-ptp13}}

\ret{Note that eq. \ref{eq:fgr-1p} is }valid  for any form of $\hat H_{b,1}$ and $\hat H_{b,2}$ and thus can be used to understand the effect of anharmonic terms in combination with appropriate quantum dynamics calculations.  In fact, the total Hamiltonian defined here, $\hat H_0+\hat H_c$, is amenable for \rc{application of various  quantum dynamics calculation methods currently available  because} $|\psi_{e,1}\rangle$ and $|\psi_{e,2}\rangle$ are diabatic states.  \rc{Therefore, direct quantum dynamics calculation beyond rate description is possible under the same quasiadiabatic approximation as well.}

\subsubsection{Closed-form expression for linearly coupled harmonic oscillator bath}
For the case where contributions of anharmonic terms are negligible, the bath Hamiltonians can be approximated as sums of harmonic oscillators as detailed in Sec. II of SI. Let us consider the simplest case where all the oscillators can be approximated as displaced ones, for which $\omega_{j,2}=\omega_{j,1}$, $\hat q_{2,j}=\hat q_{1,j}+d_j$, and $\hat p_{2,j}=\hat p_{1,j}$ in Eq. (S41) of SI.  Without losing generality, we can drop the subscript $1$ for $\hat H_{b,1}$ and $\hat \rho_{b,1}$ and introduce a bath operator $\hat B$ such that  $\hat H_{b,2}=\hat H_b+\hat B+\sum_j \omega_j^2 d_j^2/2$.
Thus, $\hat H_0$ for this case can be expressed as 
\ben
\hat H_0&\approx& \Big \{E_1 +\hat H_{b} \Big \}|\psi_{e,1} \rangle\langle \psi_{e,1}|   \nonumber \\
&+&  \Big \{ E_{2}+\hat H_{b}+\hat B \Big \} |\psi_{e,2}\rangle \langle \psi_{e,2} | ,
\een
where $E_1=E_1^0$, $E_2=E_2^0+\sum_j \omega_j^2 d_j^2/2$, $\hat H_b=\sum_j\hbar\omega_j\left (\hat b_j^\dagger \hat b_j+\frac{1}{2}\right)$, and  $\hat B=\sum_j \hbar\omega_j g_j (\hat b_j+\hat b_j^\dagger)$. In the above expressions, $\hat b_j$ and $\hat b_j^\dagger$ are usual lowering and raising operators for harmonic oscillators.  Thus, $\hat b_j=\sqrt{\omega_j/(2\hbar)} \hat x_j+i\hat p_j/\sqrt{2\hbar \omega_j}$ and $\hat b_j^\dagger =\sqrt{\omega_j/(2\hbar)} \hat x_j-i\hat p_j/\sqrt{2\hbar \omega_j}$, and $g_j=d_j\omega_j/\sqrt{2\hbar\omega_j} $.  
Then, the bath correlation function for the present case can be expressed as
\be
{\mathcal C}_{jj'}(t)={\rm Tr}_b \left \{e^{i t(\hat H_{b}+\hat B)/\hbar} \hat p_{j} e^{-i t\hat H_{b}/\hbar} \hat \rho_{b} \hat p_{j'} \right\} . \label{eq:cjj't-1}
\ee
The above bath correlation function can be evaluated explicitly as described in Appendix A, leading to the following closed form expression for the FGR rate  
\ben
k_{FGR}&=&\frac{1}{\hbar^2}\int_{-\infty}^\infty dt e^{i(E_2-E_1-\lambda)t/\hbar -{\mathcal K}(t)}\nonumber \\
&&\times \Bigg\{F(t)F^*(-t)+D(t)  \Bigg \}  \nonumber \\
&=&\frac{1}{\hbar^2}\int_{-\infty}^\infty dt e^{i(E_1-E_2+\lambda)t/\hbar -{\mathcal K}^*(t)}\nonumber \\
&&\times  \Bigg\{ F(-t)F^*(t)+D(-t)  \Bigg \} , \label{eq:fgr-2}
\een
where expressions for ${\mathcal K}(t)$, $F(t)$, and $D(t)$ are given by eqs. \ref{eq:kt-def}, \ref{eq:ft-def} and \ref{eq:dt-def} respectively.   The second equality in the above equation is obtained by either taking complex conjugate of the integrand or replacing $t$ with $-t$ of the first expression.  Thus, ${\mathcal K}^*(t)={\mathcal K}(-t)$.  We provide this latter expression since it corresponds to more conventional one and will be used for the numerical calculation.    

Equation \ref{eq:fgr-2} is the major result of the present work.  \ret{For large molecules or molecules embedded in other environments, the distribution of normal vibrational modes included in  ${\mathcal K}(t)$, $F(t)$, and $D(t)$ forms (near) continuum.  For this case, it is necessary to represent them in terms of bath spectral densities.  For ${\mathcal K}(t)$, the following spectral density is well established: 
\be 
{\mathcal J}(\omega)=\pi\hbar \sum_j \omega_j^2 \delta (\omega-\omega_j) g_j^2 . \label{eq:j-spd}
\ee
For $D(t)$ and $F(t)$, we introduce two new spectral densities as follows: }  
\ben
&&{\mathcal J}_D(\omega)=\pi\sum_j  \frac{\omega_j}{2}\delta (\omega -\omega_j) |\tilde F_{j,12}|^2  ,  \\
&&{\mathcal J}_ F(\omega)=-i\pi\sum_j  \left (\frac{\hbar\omega_j}{2}\right)^{1/2} \omega_j \delta (\omega-\omega_j) \tilde F_{j,12} g_j  . \label{eq:jf-spd}
\een
Note that ${\mathcal J}_ D(\omega)$ is real valued and positive as is ${\mathcal J}(\omega)$.  ${\mathcal J}_ F(\omega)$, which represents the sum of  couplings between displacements \rc{of normal modes and the projections of the first NDC terms onto corresponding modes}, is also real valued because $\tilde F_{j,12}$ is purely imaginary under the assumption that all the adiabatic electronic wave functions  in eq. \ref{eq:tf-j} are real valued.   However, ${\mathcal J}_ F(\omega)$  is not necessarily positive unlike ${\mathcal J}(\omega)$ and ${\mathcal J}_D(\omega)$ although it is still bounded in its magnitude as follows:
\be 
\left | {\mathcal J}_F (\omega) \right | \leq \frac{1}{2}\left ({\mathcal J}(\omega)+{\mathcal J}_D(\omega)\right) . \label{eq:jf-crit}
\ee
Then, the bath correlation functions that appear in eq. \ref{eq:fgr-2} can be expressed \rc{in terms of the spectral densities defined above as follows:}
\ben
{\mathcal K}(t)&=&\frac{1}{\pi\hbar}\int_0^\infty  d\omega\frac{{\mathcal J}(\omega)}{\omega^2} \Big (\coth(\frac{\beta\hbar \omega}{2})(1-\cos (\omega t))\nonumber \\
&&\hspace{1.in}-i\sin (\omega t)\Big ) \nonumber \\
&\equiv&{\mathcal K}_R(t)-i{\mathcal K}_I(t), \label{eq:kt-def-s}\\
D(t)&=&\frac{\hbar}{\pi}\int_0^\infty d\omega {\mathcal J}_ D(\omega) \Big (\coth(\frac{\beta\hbar \omega}{2} ) \cos (\omega t)\nonumber \\
&&\hspace{1.in} +i\sin (\omega t)\Big ) \nonumber \\
&&\equiv D_R(t)+iD_I(t) , \label{eq:dt-def-s} \\
F(t)&=&\frac{i}{\pi}\int_0^\infty d\omega \frac{{\mathcal J}_ F(\omega)}{\omega} \Big (\coth(\frac{\beta\hbar \omega}{2})(1-\cos (\omega t)) \nonumber \\
&&\hspace{1.in} -i\sin (\omega t)\Big ) \nonumber \\
&\equiv&iF_I(t)+F_R(t) . \label{eq:ft-def-s}
\een

The above representations of time correlation functions in terms of bath spectral densities have significant implications both conceptually and practically.  They retain all the necessary information on the effects of molecular vibrations and environmental responses within the linear response approximation and can be employed to some extent even beyond  the model of linearly coupled harmonic oscillator bath models.  Determination of ${\mathcal J}(\omega)$ from the energy gap correlation function\cite{jang-rmp90,cho-jcp159} of dynamics  simulation is well established and has been confirmed to work well for the modeling of absorption spectra.  For the present purpose and also for emission spectra,  \rc{${\mathcal J}(\omega)$ can be calculated through dynamics simulation for} the excited potential energy surface. \rc{For ${\mathcal J}_D(\omega)$  and ${\mathcal J}_F(\omega)$, similar approaches incorporating the information on NDC terms can also be developed}.

\rc{\section{Model calculations}
We here consider} a generic model for a molecule with one prominent vibrational frequency embedded in liquid or disordered solid environments, where low frequency vibrational modes of molecules plus environmental effects are typically modeled well in terms of Ohmic bath spectral densities.\cite{weiss,nitzan,jang-jpcb106,jang-rmp90,jang-exciton,ritschel-jcp141,zaccone-pnas118}  Given that the spectral range of the Ohmic bath is smaller than the frequency of the molecular vibration, this situation can be modeled well\cite{onuchic-jpc90,sun-jcp144-2} by the following sum of Ohmic and delta function spectral densities:  
\be
{\mathcal J}(\omega)=\pi \lambda_l\frac{\omega}{\omega_c} e^{-\omega/\omega_c}+\pi\lambda_h \omega_h \delta (\omega-\omega_h) ,  \label{eq:spd-j}\\
\ee
where $\lambda_l$ and $\lambda_h$ are components of reorganization energies due to the low frequency Ohmic part and the isolated high frequency parts. 
The same model was considered\cite{jang-jcp155-1} for the generalization of the energy gap law.

For the present work, we also need to consider the forms of bath spectral densities for ${\mathcal J}_D(\omega)$ and ${\mathcal J}_F(\omega)$ as well.  Since there are not sufficient physical or computational data for these spectral densities yet, we suppose similar forms for these as follows: 
 \ben
&&{\mathcal J}_D(\omega)=\pi D_l\frac{\omega}{\omega_c} e^{-\omega/\omega_c}+\pi D_h\omega_h \delta (\omega-\omega_h ) , \\
&&{\mathcal J}_F(\omega) =\pi F_l \frac{\omega}{\omega_c} e^{-\omega/\omega_c}+\pi F_h \omega_h\delta (\omega-\omega_h) ,
\een
where $D_l$ and $D_h$ represent squared magnitudes of NDC terms, whereas $F_l$ and $F_h$ correspond to sums of couplings between the NDC and \ret{vibronic displacement} terms \ret{over all the vibrational modes}.  These four parameters are all defined in the units of energy.   Alternatively, these can be expressed in terms of dimensionless parameters, $\eta$, $s_h$, $\eta_{_D}$, $s_{_D}$, $\eta_{_F}$, and $s_{_F}$, as indicated in Table \ref{table-parameters}.  Since the high frequency delta function part originates from a specific molecular vibration, the magnitude of $s_{_F}$ is determined given the values of $s_h$ and $s_{_D}$ such that $s_{_F}^2=s_hs_{_D}$. On the other hand,  the value of $F_l$ is not fully determined except the condition that $|F_l|\leq D_l/2$, which results from eq. \ref{eq:jf-crit}. 
\begin{table}

\caption{Relationship between parameters defining bath spectral densities.}
\label{table-parameters}
\begin{tabular}{cccccc}
\hline
\hline
\makebox[.5in]{$\lambda_l$}&\makebox[.5in]{$\lambda_h$}&\makebox[.5in]{$D_l$}&\makebox[.5in]{$D_h$}&\makebox[.5in]{$F_l$}&\makebox[.5in]{$F_h$} \\
\hline
$\eta\hbar\omega_c$ &$s_h\hbar\omega_h$ & $\eta_{_D}\hbar\omega_c$ & $s_{_D}\hbar\omega_h$ & $\eta_{_F}\hbar\omega_c$ &$s_{_F}\hbar\omega_h$  \\
\hline
\hline
\end{tabular}
\end{table}
\ \vspace{.4in}\\

For the spectral density of eq. \ref{eq:spd-j}, the real and imaginary parts of ${\mathcal K}(t)$, as defined through the last line of eq. \ref{eq:kt-def}, can be expressed as
\ben
&&{\mathcal K}_R(t)=\frac{1}{\pi\hbar}\int_0^\infty d\omega \frac{{\mathcal J}(\omega)}{\omega^2} \coth \left (\frac{\beta\hbar\omega}{2}\right) \left (1-\cos(\omega t)\right) \nonumber \\
&&\approx\frac{\lambda_l}{\hbar\omega_c} \left \{ \frac{1}{2} \ln (1+\tau_0^2 )+\ln (1+\tau_1^2) +\ln (1+\tau_2^2) \right . \nonumber \\
&&\hspace{.1in}\left . + \frac{2(1+5\theta/2)}{\theta} \left [ \tau_{5/2}\tan^{-1}(\tau_{5/2}) -\frac{1}{2} \ln (1+\tau_{5/2}^2) \right ]  \right \} \nonumber \\
&&\hspace{.1in}+\frac{\lambda_h}{\hbar\omega_h} \coth \left(\frac{\beta\hbar\omega_h}{2}\right)\left (1-\cos (\omega_h t) \right ) , \label{eq:kr-t} \\
&&{\mathcal K}_I(t)=\frac{1}{\pi\hbar}\int_0^\infty d\omega \frac{{\mathcal J}(\omega)}{\omega^2}\sin (\omega t)  \nonumber \\
&&\hspace{.35in}=\frac{\lambda_l }{\hbar\omega_c} \tan^{-1} (\tau_0)+\frac{\lambda_h}{\hbar\omega_h}\sin (\omega_h t)  , \label{eq:ki-t-m}
\een
where $\theta=\beta\hbar\omega_c$ and $\tau_n=\omega_c t/(1+n\theta)$.  The second equality in eq. \ref{eq:kr-t} is based on the following approximation:\cite{jang-jpcb106}
\be
\coth\left (\frac{\beta\hbar\omega}{2}\right )\approx 1+2 e^{-\beta\hbar\omega}+2 e^{-2\beta\hbar\omega}+\frac{2}{\beta\hbar\omega} e^{-5\beta\hbar\omega/2} ,  \label{eq:coth-app}
\ee
and we have used an explicit expression for $\int_0^{\tau_{5/2}} d\tau' \tan^{-1} (\tau')$.
On the other hand,  the real and imaginary parts of $D(t)$, which are defined by the last line of eq. \ref{eq:dt-def}, are expressed as
\ben
D_R(t)&=&\frac{\hbar}{\pi} \int_0^\infty d\omega {\mathcal J}_D(\omega) \coth \left (\frac{\beta\hbar\omega}{2}\right) \cos(\omega t) \nonumber \\
&\approx&\hbar \omega_c D_l \left \{\frac{1-\tau_0^2}{(1+\tau_0^2)^2} +\frac{2}{(1+\theta)^2}\frac{1-\tau_1^2}{(1+\tau_1^2)^2} \right . \nonumber \\
&&\hspace{.4in}+\frac{2}{(1+2\theta)^2}\frac{1-\tau_2^2}{(1+\tau_2^2)^2} \nonumber \\
&&\hspace{.4in}\left .+\frac{2}{\theta (1+5\theta/2)}\frac{1}{(1+\tau_{5/2}^2)}\right\}  \nonumber \\
&&+\hbar \omega_h D_h \coth \left (\frac{\beta\hbar\omega_h}{2}\right) \cos(\omega_h t) , \label{eq:dr-t-mod}\\
D_I(t)&=&\frac{\hbar}{\pi} \int_0^\infty d\omega {\mathcal J}_D(\omega) \sin (\omega t) \nonumber \\
&=&\hbar \omega_c D_l \frac{2\tau_0}{(1+\tau_0^2)^2}+\hbar \omega_h D_h \sin (\omega_h t) .
\een
Finally, the real and imaginary parts of $F(t)$, which are defined by the last line of eq. \ref{eq:ft-def}, have the following expressions: 
\ben
F_R(t)&=&F_l \frac{\tau_0}{(1+\tau_0^2)}+F_h\sin(\omega_h t) , \\
F_I(t)&\approx &F_l \left (\frac{\tau_0^2}{(1+\tau_0^2)}+\frac{2}{(1+\theta)}\frac{\tau_1^2}{(1+\tau_1^2)} \right .  \nonumber \\
&&\left .+\frac{2}{(1+2\theta)}\frac{\tau_2^2}{(1+\tau_2^2)}+\frac{1}{\theta}\ln (1+\tau_{5/2}^2 )\right ) \nonumber \\
&&+F_h \coth \left (\frac{\beta\hbar\omega_h}{2}\right)\left (1-\cos (\omega_h t)\right ) . \label{eq:fi-t-mod}
\een
Note that the second approximate equalities in eqs. \ref{eq:dr-t-mod} and \ref{eq:fi-t-mod} are also based on the approximation of eq.  \ref{eq:coth-app}.

For numerical calculations, we considered two values of $\omega_h/\omega_c=5$ and $15$, which respectively represent moderate and high frequency vibrational frequencies.  \ret{Given that $\hbar\omega_c\approx 200\ {\rm cm^{-1}} $, the thermal energy at room temperature, the two choices of $\omega_h$ correspond to about $1,000$ and $3,000\ {\rm cm^{-1}}$ for the vibrational frequencies.}  For each choice, we  then considered three different cases of other parameters such that the sum of $D_l$ and $D_h$ remains the same.  The case A represents the situation where NDC terms solely originate from the isolated vibrational mode.   Cases B and C represent examples where Ohmic bath parts make dominant contributions to NDC terms.  In the presence of two different sources for NDC terms, the sign of $s_{_F}$ relative to $\eta_{_F}$ \rc{has different effects on rates}.  Cases B and C respectively correspond to positive and negative signs of $s{_{F}}$, \rc{while having the same values for all other parameters}.  In addition to $k_{FGR}$ calculated exactly according to eq. \ref{eq:fgr-2}, we also calculated rates based on the following Condon approximation:    
\be
k_{Con.}=\frac{1}{\hbar^2} D_R(0)\int_{-\infty}^\infty dt e^{i(E_1-E_2+\lambda)t/\hbar -{\mathcal K}^*(t)} . \label{eq:k-app}
\ee
The above rate expression employs NDC terms evaluated at $t=0$ as effective Condon-coupling terms.
\begin{table}
\caption{Table of model parameters for spectral densities.  Two other parameters are set to $k_BT/(\hbar\omega_c)=1$ and $\eta=1$. Note that $s_{_F}^2=s_hs_{_D}$ and the result is independent of the sign of $s_{_F}$ for the case $\eta_{_F}=0$. }
\label{table-parameter}
\begin{tabular}{c|cccccc} \hline \hline
{\bf Case} &\makebox[.5in]{$\omega_h/\omega_c$ }&\makebox[.4in]{ $s_h$}& \makebox[.4in]{$\eta_{_D}$} &\makebox[.4in]{$\eta_{_F}$}  &\makebox[.4in]{$s_{_D}$}&\makebox[.5in]{$s_{_F}$}\\
\hline
{\bf I-A} &5&1&0&0&4&$2$\\
{\bf I-B}&5&1&15&2&1& $1$\\
{\bf I-C}&5&1&15&2&1& $-1$\\
\hline
{\bf II-A} &15&0.2&0&0&1& $\sqrt{0.2}$\\
{\bf II-B}&15&0.2&12&2&0.2& $0.2$\\
{\bf II-C}&15&0.2&12&2&0.2& $-0.2$\\
\hline
\hline
\end{tabular}
\end{table} 
\ \vspace{.5in}\\

\begin{figure}
\ \vspace{.4in}\\
\includegraphics[width=3.in]{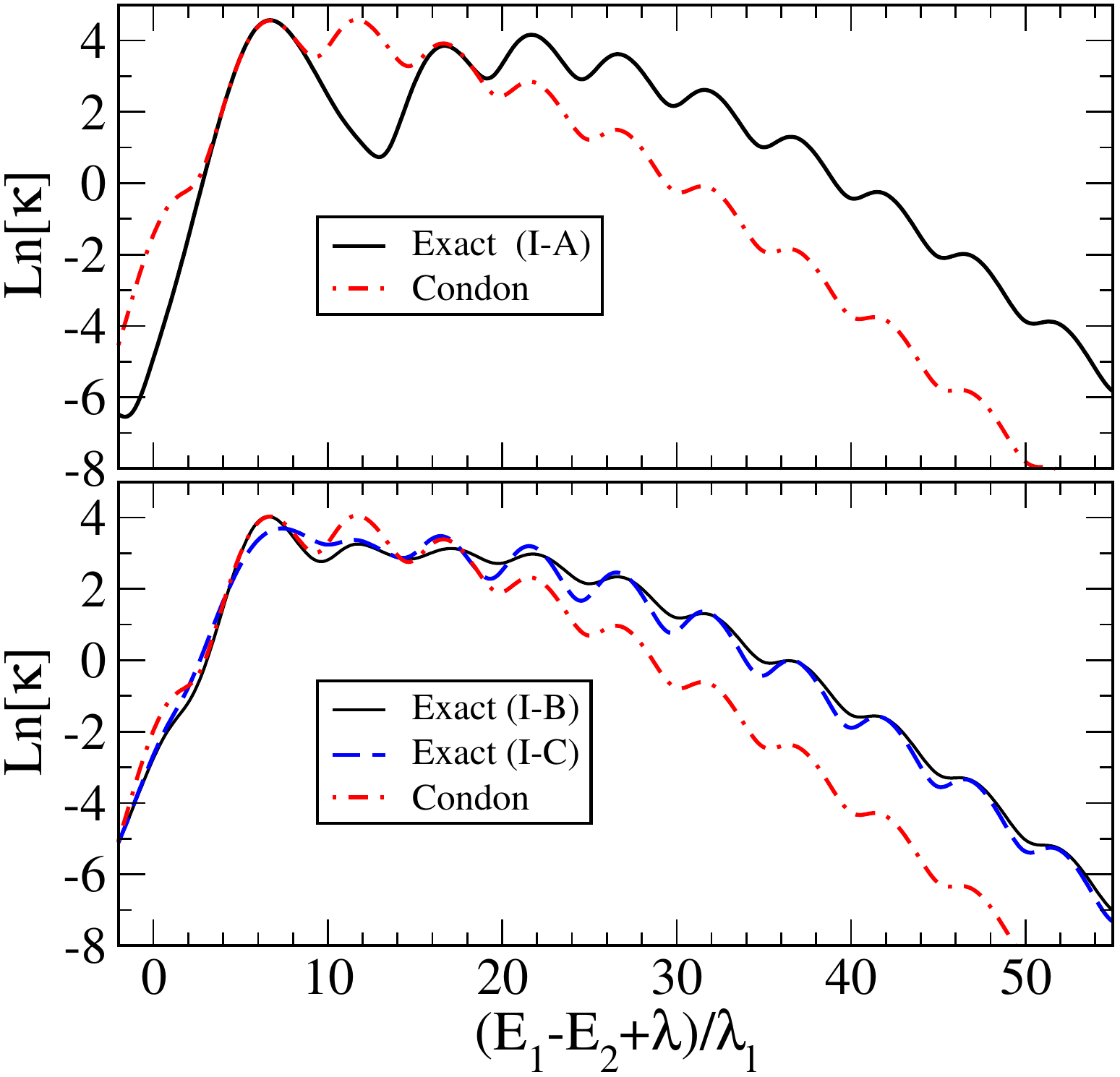}
\caption{Natural logarithms of scaled rates $\kappa=\sqrt{k_BT \lambda/\pi} k_{FGR}$ (in the units where $k_B=\hbar=\omega_c=1$) for case I-A (upper panel) and  cases I-B and I-C (lower panel), are compared with reference results of a Condon approximation, eq. \ref{eq:k-app}, for the same cases.  }
\label{figure1}
\end{figure}

Figure \ref{figure1} shows results for cases I-A,B, and C of Table \ref{table-parameter}.  The upper panel shows results for case I-A, and the lower panel compares the two results for cases I-B and I-C.  The Condon approximations for these cases are the same, while slightly smaller than that for case I-A.  Oscillatory behavior of rates with respect to the energy gap reflects the vibrational progression due to the single vibrational mode.  It is clear that non-Condon effects due to momentum terms result in significant enhancement of rates \rc{as the energy gap increases}, when compared with the Condon approximation \rc{given by eq. \ref{eq:k-app}} in all cases.  

Comparison of upper and lower panels of Fig. \ref{figure1} in the large energy gap limit shows that the rates for the case A are consistently larger than those for cases B and C, which means that NDC terms coming from the single high frequency vibrational mode are more effective in enhancing the rate compared with those from the lower frequency Ohmic part.  Nonetheless, except for small or negative values of $E_1-E_2$, for which the validity of the quasi-adiabatic approximation may not be fully justified, the difference between the results for upper and lower panels are relatively small.  
The lower panel of Fig. \ref{figure1} shows that the case with negative $s_{_F}$ results in smaller rates than those for positive $s_{_F}$, as expected, but the difference between the two are relatively small as well. Considering  all of the observations made for Fig. 1, we conclude that ${\mathcal K}(t)$ has the most dominant effect, followed by $D(t)$.  The role of $F(t)$ is relatively minor compared to others.

\begin{figure}
\ \vspace{.4in}\\
\includegraphics[width=3.in]{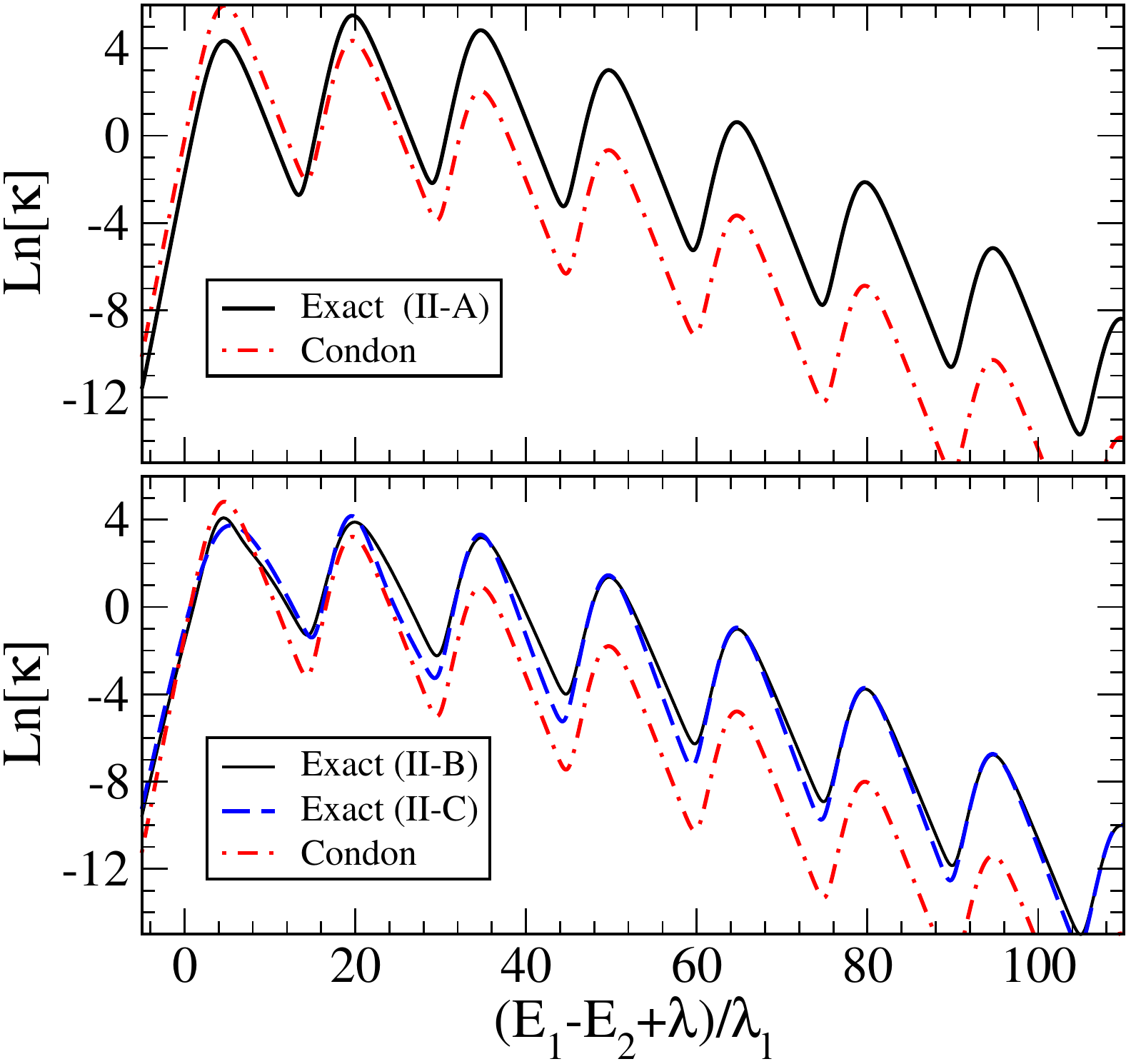}
\caption{Natural logarithms of scaled rates $\kappa=\sqrt{k_BT \lambda/\pi} k_{FGR}$ (in the units where $k_B=\hbar=\omega_c=1$) for case II-A (upper panel) and cases II-B and II-C (lower panel),  are compared with  reference results of a Condon approximation, eq. \ref{eq:k-app}, for the same cases.}
\label{figure2}
\end{figure}

Figure \ref{figure2}  shows results for the cases II-A, B, and C  of Table \ref{table-parameter}, for which the frequency of the vibrational mode is three times larger than those for Fig. \ref{figure1}.  Although the reorganization energy of the single vibrational frequency in this case is smaller than that for Fig. \ref{figure1}, more pronounced vibrational progression effect can be seen.   Otherwise, qualitative results are similar to those for Fig. \ref{figure1}.   

\begin{figure}
\ \vspace{.4in}\\
\includegraphics[width=3.in]{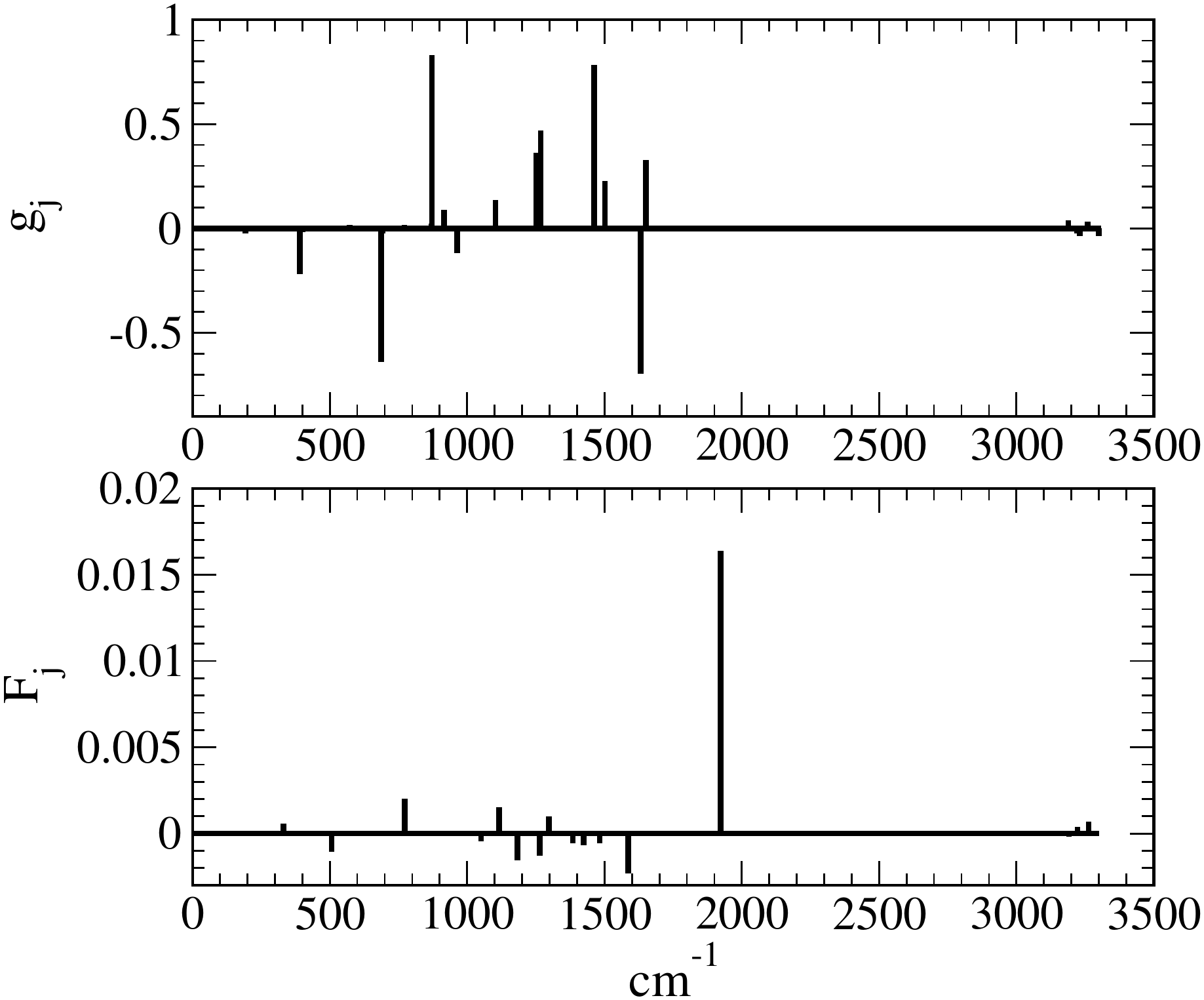} \\
\caption{Histograms of $g_j$, which is dimensionless, and ${\rm Im}\ \tilde F_j$ (in atomic units) versus the wavenumber of normal modes for azulene.}
\label{figure3}
\end{figure}
 
\rc{\section{Application to azulene}
We conducted calculations} for an actual molecule, azulene, employing the time dependent density functional method  with M06-2X functional\cite{zhao-tca120} and  6-31+G* basis.   We optimized the molecule in its first excited state in the gas phase and calculated Huang-Rhys (HR) factors and the first NDC terms corresponding to the ${\rm S_1 \rightarrow S_0}$ transition for all the vibrational modes, for an appropriate Eckart frame identified through a well established numerical procedure.\cite{rhee-jcp113} Figure \ref{figure3} shows histograms of HR factors and NDC terms.  It is interesting to note that the contribution of CH stretching modes on both HR and first NDC terms are very small for this transition.  It is also interesting to note that the contribution to NDC term is dominated by one vibrational frequency at $1,924 \ {\rm cm^{-1}}$, whereas the HR terms are contributed by a group of vibrational modes with smaller frequencies.  Due to \rc{very small couplings} between the two, it is expected that  the contribution of $ F(t)$ to the rate is insignificant.

\begin{figure}
\ \vspace{.4in}\\
\includegraphics[width=3.in]{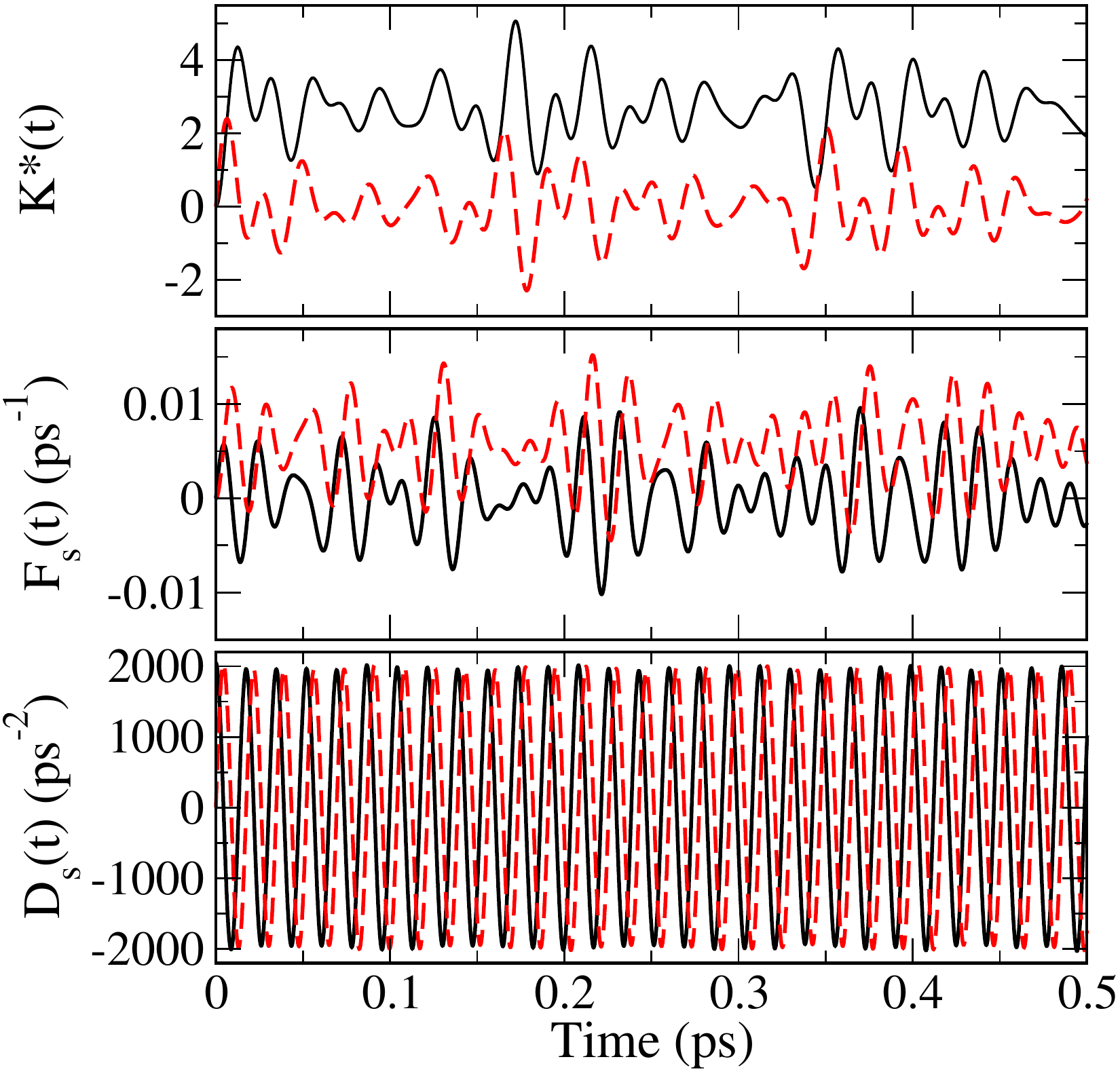}
\caption{Plots of ${\mathcal K}^*(t)$, $F_s(t)=F(t)/\hbar$ (in the unit of ${\rm ps^{-1}}$), and $D_s(t)=D(t)/\hbar^2$ (in the unit of ${\rm ps^{-2}}$) versus time (in the unit of ${\rm ps}$) for azulene. Black solid lines are real parts and red dashed lines are imaginary parts.}  
\label{figure4}
\end{figure}

Figure \ref{figure4} shows real and imaginary parts of ${\mathcal K}^*(t)$ (which is dimensionless), and $F(t)/\hbar$ and $D(t)/\hbar^2$ (in the units of ${\rm ps^{-1}}$ and ${\rm ps^{-2}}$ respectively) due to all the vibrational modes determined for an isolated azulene at ${\rm 300\ K}$.  \rc{These were calculated directly from eqs. \ref{eq:kt-def}, \ref{eq:ft-def}, and \ref{eq:dt-def}. This amounts to calculating eqs. \ref{eq:kt-def-s}-\ref{eq:ft-def-s} employing an exact delta function for the three bath spectral densities defined by eqs. \ref{eq:j-spd}-\ref{eq:jf-spd}. Comparison of relative magnitudes of $F(t)$ and $D(t)$ in Fig. \ref{figure4} clearly shows that the latter can practically account for all contributions of NDC terms for this molecule.  }

For the calculation of rates, we have also included the effects of environments in two different ways.  One was to add the contribution of the Ohmic bath, the first term in eq. \ref{eq:spd-j}, only to ${\mathcal K}(t)$, \rc{which amounts to adding additional ${\mathcal K}_R(t)$ and ${\mathcal K}_I(t)$ given by eqs \ref{eq:kr-t} and \ref{eq:ki-t-m} with $\lambda_h=0$.}  This represents the situation where the environmental dynamics can be represented by an Ohmic bath with spectral range much narrower than the major molecular vibrational frequencies, whereas the modes contributing to derivative couplings remain harmonic.  

\rc{The other way to include environmental effects was to }represent each delta function in the definition of spectral densities, eqs. \ref{eq:j-spd}-\ref{eq:jf-spd}, with the following normalized Brownian oscillator (BO) model:\cite{garg-jcp83,mukamel}
\be
\delta (\omega-\omega_j)\approx J_{\gamma}(\omega;\omega_j)=\frac{A_j}{\gamma^2}\frac{\omega}{\left [ (\omega^2-\omega_j^2)^2/\gamma^4+4(\omega/\gamma)^2\right]} , \label{eq:bo-spd}
\ee
where $A_j$ is the normalization constant determined such that integration of $J_{\gamma}(\omega;\omega_j)$ over [0,$\infty$] is unity and $\gamma$ is a parameter representing the friction of environment. Explicit expression for $A_j$ is provided by eq. S51 of SI. \rc{The above BO model behaves like a Lorentzian near $\omega_j$, while being free of unphysical limiting behavior\cite{sluis-pra43,elyutin-jl88} of the latter that is responsible for numerical ambiguities or artifacts.  More specifically, unlike the Lorentzian,  the above BO model } approaches $\omega=0+$ linearly  and decays as $\omega^{-3}$ for large $\omega$. \rc{These limiting properties make all three bath correlation functions of present work well defined and amenable for straightforward numerical integrations.  More details of numerical procedure are provided in SI.}   In addition, the above BO model represents \rc{an actual physical situation} where each vibrational mode is weakly coupled to a broad Ohmic bath and can also be used to some extent to represent the broadening due to the anharmonicity\cite{cho-jcp159} of the bath degrees of freedom. 

\begin{figure}
\ \vspace{.4in}\\
\includegraphics[width=3.in]{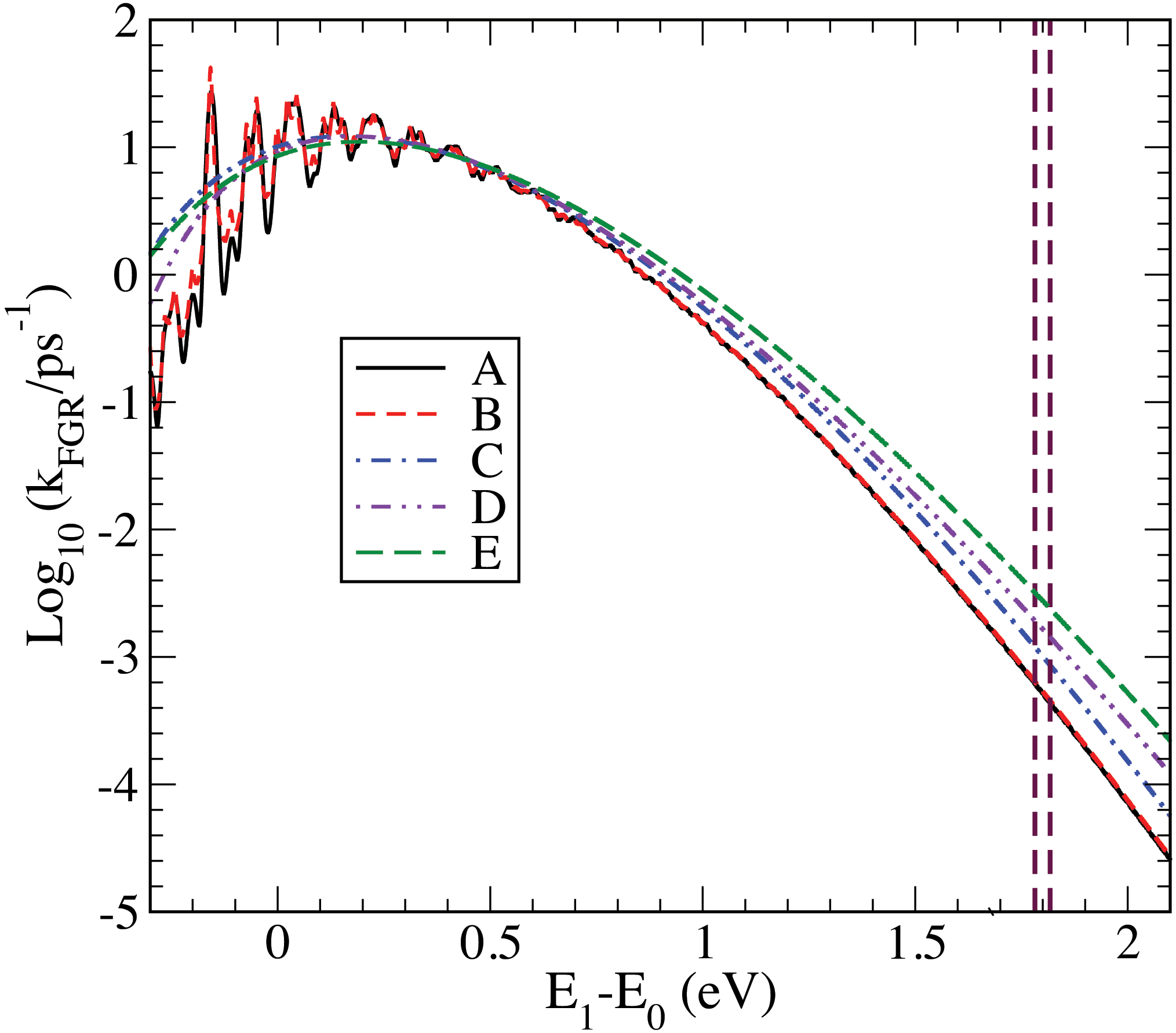}
\caption{\ret{FGR rates of nonradiative decay from ${\rm S_1}$ to ${\rm S_0}$ versus the energy gap $E_1-E_0$ of azulene for five different bath model parameters provided in Table \ref{table-bath}.  The two vertical lines represent the minimum and maximum values of the energy gap values in solution.}}
\label{figure5}
\end{figure}

\begin{table}
\caption{\ret{Table of bath model parameters, $\eta$ and $\tilde \nu_c =\omega_c/(2\pi c)$ for the Ohmic bath spectral density defined by the first term of eq. \ref{eq:spd-j} and Table \ref{table-parameters}, and  $\tilde \gamma =\gamma/(2\pi c)$, where $\gamma$ is defined by eq. \ref{eq:bo-spd}.  Rates calculated at two values of the energy gap are shown.  $k_{min}$ is for $E_{1}-E_0=1.8176\ {\rm eV}$ and $k_{max}$ is for $E_{1}-E_0=1.7816\ {\rm eV}$.} }
\label{table-bath}
\begin{tabular}{c|ccccc} \hline \hline
{\bf Bath} &\makebox[.2in]{ $\eta$}& \makebox[.6in]{$\tilde \nu_c\ ({\rm cm^{-1}})$ }&\makebox[.6in]{$\tilde \gamma\  ({\rm cm^{-1}})$} &\makebox[.7in]{$k_{min}\ ({\rm ns^{-1}})$}  &\makebox[.7in]{$k_{max}\ ({\rm ns^{-1}}) $}\\
\hline
{\bf A} &1&10&0&0.439&0.617\\
{\bf B}&0&&10&0.453&0.639\\
{\bf C}&10&200&0&0.861&1.20\\
{\bf D} &0&&400&1.43&1.93\\
{\bf E}&10&200&400&2.40&3.20\\
\hline
\hline
\end{tabular}
\end{table} 
\ \vspace{.5in}\\
\begin{figure}
\ \vspace{.4in}\\

\includegraphics[width=3.in]{Figures/Figure6.eps}
\caption{\rc{Plots of ${\mathcal K}_R(t)$ versus time for the five different cases A-E of Table \ref{table-bath} for azulene, shown as black lines.  The data without bath (which was provided in Fig. \ref{figure4}) are shown as red dashed lines as a  reference.}}  
\label{figure6}
\end{figure}

Figure \ref{figure5} shows rates calculated according to eq. \ref{eq:fgr-2} at ${\rm 300\ K}$ for \rc{five different choices  as listed in Table \ref{table-bath}, where the values of parameters, $\eta$, $\tilde \nu_c=\omega_c/(2\pi c)$, and $\tilde \gamma=\gamma /(2\pi c)$ are shown.   Figure \ref{figure6} provides ${\mathcal K}_R(t)$s for the five cases calculated during the short initial time, in comparison with that in the absence of bath (Same data for longer time are shown in top panel of  Fig. \ref{figure4}.)  Similar figures for all other correlation functions are provided in SI.}   
Two vertical dashed lines in \rc{Fig. \ref{figure5}} represent the lower and upper bounds of experimental energy gaps\cite{wagner-jcp98} determined in different solution phase at room temperature and correspond respectively to $1.7816\ {\rm eV}$ and $1.8176\ {\rm eV}$. Table \ref{table-bath} also lists  minimum and maximum values of FGR rates at the two values of the energy gap.  
 
 Bath models A and B correspond to the cases where either the explicit Ohmic bath or the friction coefficient in the BO model is much smaller than other frequencies.  Good agreement of the two, especially in the large energy gap limit,  suggests that they are close to intrinsic rates without environmental effects. \rc{The similarity of ${\mathcal K}_R(t)$ in the short time limit for these two cases, as shown in Fig. \ref{figure6}, confirms this as well.} Bath models \rc{C-E} represent intermediate couplings to environments, \rc{for which ${\mathcal K}_R(t)$ starts increasing quickly already around $t=0.1\ {\rm ps}$, as shown in Fig. \ref{figure6}.}   While \rc{model C} represents the effects of low frequency Ohmic bath, \rc{model D} represents significant broadening of each vibrational mode due to couplings to other environmental modes.  Model E corresponds to the combination of the two.       

It is interesting to note that our highest theoretical estimates are comparable to that  by Niu {\it et al.}\cite{niu-jpca114} under a similar assumption.  
The upward trend in the rate we observe for larger reorganization energy of the environment suggests that stronger environmental effects on both HR and NDC terms can enhance the rate.  Additional contributions of Duschinsky effects\cite{niu-jpca114} \rc{may} increase the rate.  Anharmonic effects can also enhance the rate significantly.\cite{wang-jcp154}  On the other hand, a very recent theoretical work by Landi {\it et al.}\cite{landi-jcp160} reported new theoretical estimates, $62\ {\rm ns^{-1}}$  and $71\ {\rm ns^{-1}}$, based on direct quantum dynamics calculation and the generating function approach\cite{kubo-ptp13} within the FGR rate respectively for a similar harmonic approximation for all the vibrational modes.  It was also reported\cite{landi-jcp160} that the net effect of the Duschinsky effect is not significant for the value of the energy gap \rc{as opposed to the assessment by Shuai and coworkers.\cite{niu-jpca114} As yet, although most advanced, the computational data by Landi {\it et al.}\cite{landi-jcp160} seem to require further validation to serve as benchmark data. For example, the convergence of the direct quantum dynamics calculation data needs to be examined carefully.  Further clarification of the effects of  window or apodization function within the generating function seems also necessary. Although detailed and clear theoretical analyses within the generating function approach are available for the simple cases\cite{nitzan-tca29,nitzan-tca30,kuhn-cp33,freed-acr11} of Lorentzian or Gaussian broadening mechanisms, issues\cite{kuhn-cp33} related to the  behavior resulting from small and large frequency limits of the Lorentzian function have not been fully resolved.  While Gaussian broadening results in consistent rates,\cite{kuhn-cp33}  its physical origin except for two cases, an ensemble of static disorder and strongly coupled environments, are not clearly understood yet.  On the other hand, the rates provided in this section are based on more clear physical assumption and are free of the effects of any extra window or apodization function.} 

While the theoretical estimates for the nonradiative rate for ${\rm S_1}\rightarrow {\rm S_0}$ are relatively large  when compared to other organic molecules of similar size, they are still at least two  orders of magnitude smaller than experimental estimates, which were reported\cite{ippen-cpl46,shank-cpl57} to be about $0.5\ {\rm ps^{-1}}$ but have since been revised\cite{schwarzer-bbpc95,wagner-jcp98} later to be about  $1\ {\rm ps^{-1}}$ or faster.  This discrepancy is significant and points to the possibility of other mechanisms.   Indeed, there have been theoretical and experimental works\cite{bearpark-jacs118,diau-jcp110,wurzer-cpl299,amatatsu-jcp125,dunlop-jacs145} suggesting that the major mechanism for fast nonradiative  decay of azulene from ${\rm S_1}$ is via conical intersection.  While the initial theoretical assessment of the mechanism\cite{bearpark-jacs118} was revised later,\cite{amatatsu-jcp125} additional spectroscopic measurements\cite{diau-jcp110,wurzer-cpl299,dunlop-jacs145} and computational study\cite{dunlop-jacs145} are still in strong support of decay through a conical intersection point. However, direct and quantitative demonstration of this mechanism through advanced dynamics calculations still needs to be made.  \rc{In particular, the relative slowness of the process indicates that some activation process might be involved even if the decay occurs through a conical intersection.  However, even if  this scenario were true, why the rate is independent of temperature is not well understood.  Thus, unambiguous and quantitative clarification of the ${\rm S_1\rightarrow S_0}$ nonradiative decay of azulene remains a major theoretical challenge that still needs to be resolved. } 

\section{Conclusion}
Starting from a general expression for the molecular Hamiltonian in the adiabatic electronic states and nuclear position states, we have considered NDC terms carefully and provided a full formal expression for the FGR rate expression for nonadiabatic transitions between adiabatic states.  We then showed that a general expression for the FGR rate, eq. \ref{eq:fgr-1p}, can be obtained within a quasi-adiabatic approximation, which employs crude adiabatic electronic states determined at the minimum of  the initial adiabatic electronic state.  This expression turns out  to be equivalent to the general expression (without assuming a promoting mode) being used by Shuai and coworkers.\cite{niu-jpca114} 
For the case where all the nuclear dynamics are modeled as displaced harmonic oscillators, we then derived a closed form expression for the FGR rate,  eq. \ref{eq:fgr-2}, which can be used for general distribution of vibrational modes and arbitrary couplings between Franck-Condon modes and NDC terms, while still accounting for non-Condon effect on the rate due to nuclear momenta terms exactly.  This expression also clarifies that \rc{the nonadiabatic FGR rate requires the information of} not only full vibrational modes coupled to electronic transitions but projections of derivative coupling terms along different vibrational modes.  All of these can be collectively specified by three different bath spectral densities given by eqs. \ref{eq:j-spd}-\ref{eq:jf-spd}.

We have conducted model calculations for cases where the bath spectral density consists of a low frequency Ohmic bath (with an exponential cutoff) and a single high frequency vibrational mode.  Results of calculation for sets of parameters in Table \ref{table-parameter}  demonstrate nontrivial non-Condon effects due to NDC terms.   Overall, there is consistent enhancement of rate,  with NDC terms due to higher frequencies making more significant contributions for larger energy gap between the donor and acceptor in general. 

Application of our theoretical expression to the nonradiative decay rate of azulene clarifies major factors contributing to the decay of the first excited singlet state via direct nonadiabatic coupling.  
\ret{In particular, it is shown that the broadening of spectral density due to environmental effects or anharmonic contribution can enhance the rate significantly.  As yet, we estimate that such enhancement is not sufficient to account for fast experimental nonradiative decay rates of azulene.  Thus, our calculation serves as } indirect support for the decay through conical intersection as major route for fast picosecond time scale decay observed experimentally.  Our rate expression makes it possible to clarify this issue further in combination with more extensive dynamics calculations that allow more satisfactory determination of bath spectral densities.

Results of the present paper offer new insights into rate processes due NDC terms such as nonradiative decay of near infrared and short-wave infrared dye molecules\cite{friedman-chem7,erker-jacs144} that follow the energy gap law.\cite{englman-mp18,jang-jcp155-1}  A recent work\cite{ramos-jpcl15} demonstrated the importance of NDC terms projected onto all vibrational frequencies of molecules, but the detailed contribution of non-Condon effects has not been clarified yet.  New theoretical expressions and model calculations provided here will help determine such effects quantitatively with input from additional computational data \rc{providing all the relevant spectral densities}.    

The major focus of the present work was nonadiabatic transitions that can be described well by a FGR rate equation \rc{and to derive} a closed form rate expression for the generic case of displaced harmonic oscillator baths. However,  further improvement accounting for Duschinsky effects and anharmonicity is feasible and will be the subject of future efforts. In addition, the validity of the molecular Hamiltonian obtained within the quasi-adiabatic approximation, eqs. \ref{eq:h-ad-1}-\ref{eq:hc-1} (or with eq. \ref{eq:hc-2} instead) is not necessarily limited by the assumption of rate behavior.  In this sense, applications of various quantum dynamics methods\cite{tully-jcp137,wu-jpcl15-1,he-jpcl15,liang-jcp149,makri-arpc50,huo-jcp137,makri-jcp146,kelly-jcp139,mandal-jctc14,beck-pr324,manthe-jcp128,burghardt-jcp129,wang-jcp146,breuer-ap291,shi-jcp119-2,shi-jcp120,ishizaki-jcp130-2,tanimura-jpsj75,tanimura-jcp142,zheng-pc24,devega-rmp89,lyu-jctc19,hsieh-jpcc123,jang-exciton,yang-jcp137,jang-jcp157} to these Hamiltonians would be interesting and can offer various benchmarks for the rate description.  This is feasible, with extension or test of known quantum dynamics methods for the types of coupling terms given by eq. \ref{eq:hc-1} or \ref{eq:hc-2}.  More general expressions provided in SI can be used for cases of multiple adiabatic states.  Given that quasi-adiabatic approximation in such cases remains valid, extension of the current FGR rate for multiple states is feasible, which again can be tested against advanced quantum dynamics calculations.  Finally, the general formalism developed for  this work also serves as a solid framework to go beyond the quasiadiabatic approximation.
\acknowledgements
S.J.J. acknowledges primary support from the US Department of Energy, Office of Sciences, Office of Basic Energy Sciences (DE-SC0021413), partial support during the initial stage of this project from the National Science Foundation (CHE-1900170), and travel support from the PSC-CUNY grant (\#67161 -0055).   Y.M.R. and B.K.M. acknowledge support from the National Research Foundation (NRF) of Korea (Grant No. 2020R1A5A1019141) and some assistance from Jiseung Kang and Hyoju Choi for electronic structure calculation of azulene.  This project initiated during sabbatical stay of S.J.J. at Korea Advanced Institute of Science and Technology (KAIST) and  Korea Institute for Advanced Study (KIAS).   S.J.J. thanks support from the KAIX program at KAIST and KIAS Scholar program for the support of sabbatical visit. 

\appendix
\section{Appendix: Calculation of bath correlation functions involving momenta}
\subsection{Preliminary calculations}
Consider the following harmonic oscillator bath Hamiltonian and a bath term linear in position: 
\ben
&&\hat H_{b}=\hbar\omega (\hat b^\dagger \hat b+\frac{1}{2}) ,\\
&&\hat B=\hbar\omega g(\hat b+\hat b^\dagger) .
\een
For the equilibrium density $\hat \rho_{b}=e^{-\beta\hat H_{b}}/Tr\left\{e^{-\beta\hat H_{b}}\right\}$, let us first consider the following well-known time correlation function:
\be
C(t)=Tr_{b} \left\{ e^{it (\hat H_{b}+\hat B)/\hbar} e^{-it \hat H_{b}/\hbar} \hat \rho_{b}\right\} .
\ee
Then, using $1=e^{- \hat S}e^{\hat S}$ with $\hat S=g(\hat b^\dagger -\hat b)$,
\ben
C(t)&=&Tr_{b} \left\{ e^{-\hat S} e^{\hat S}e^{it (\hat H_{b}+\hat B)/\hbar} e^{-\hat S}e^{\hat S}e^{-it \hat H_{b}/\hbar} \hat \rho_{b}\right\} \nonumber \\
&&=Tr_{b} \left\{ e^{-\hat S} e^{it \hat H_{b}/\hbar} e^{\hat S}e^{-it \hat H_{b}/\hbar} \hat \rho_{b}\right\} e^{-it \lambda/\hbar} \nonumber \\
&&=Tr_{b} \left\{ e^{-\hat S} e^{\hat S(t)} \hat \rho_{b}\right\} e^{-it \lambda/\hbar} , \label{eq:trbj-1}
\een
where $\lambda=\hbar\omega g^2$ and 
\be
\hat S(t)=e^{it\hat H_{b}/\hbar} \hat S e^{-it\hat H_{b}/\hbar}= g(\hat b^\dagger e^{i\omega t}-\hat b e^{-i\omega t}) .
\ee
The trace in eq. \ref{eq:trbj-1} can be evaluated as follows:
\ben
&&Tr_{b} \left\{ e^{-\hat S} e^{\hat S(t)} \hat \rho_{b}\right\} =Tr_{b} \left\{ e^{-\hat S+\hat S(t)} \hat \rho_{b}\right\}e^{-[\hat S,\hat S(t)]/2} \nonumber \\
&&=Tr_{b} \left\{ e^{g\hat b (1-e^{-i\omega t})-g\hat b^\dagger (1-e^{i\omega t} )} \hat \rho_{b}\right\}e^{ig^2\sin (\omega t)} \nonumber \\
&&=Tr_{b} \left\{ e^{g\hat b (1-e^{-i\omega t})}e^{-g\hat b^\dagger (1-e^{i\omega t} )} \hat \rho_{b}\right\}\nonumber \\
&&\hspace{1in}\times e^{g^2|(1-e^{-i\omega t}) |^2/2+ig^2\sin (\omega t)} \nonumber \\
&&=e^{-g^2|1-e^{-i\omega t}|^2/(1-e^{-\beta\hbar\omega })} e^{g^2|1-e^{-i\omega t} |^2/2}e^{ig^2\sin (\omega t)} \nonumber \\
&&=e^{-{\mathcal K}_g(t)}  , \label{eq:sst-trace}
\een
where 
\be
{\mathcal K}_g(t)=g^2[\coth(\frac{\beta\hbar\omega}{2})(1-\cos (\omega t))-i\sin (\omega t)] \label{eq:kgt}
\ee 
Therefore, 
\be 
C(t)=e^{-{\mathcal K}_g(t)-i\lambda t/\hbar}  . \label{eq:ct}
\ee

Now, consider the following time correlation function:
\be
C_p^{(1)}(t)=Tr_{b} \left\{ e^{it (\hat H_{b}+\hat B)/\hbar} \hat p e^{-it \hat H_{b}/\hbar} \hat \rho_{b}\right\} .
\ee
Following a procedure similar to Eq. (\ref{eq:trbj-1}), we find that 
\ben
C_p^{(1)}(t)&=&Tr_{b} \left\{ e^{-\hat S} e^{\hat S}e^{it (\hat H_{b}+\hat B)/\hbar} e^{-\hat S} e^{\hat S}\hat p e^{-it \hat H_{b}/\hbar} \hat \rho_{b}\right\} \nonumber \\
&=&Tr_{b} \left\{ e^{-\hat S} e^{it \hat H_{b}/\hbar} e^{\hat S}\hat p e^{-it \hat H_{b}/\hbar} \hat \rho_{b}\right\} e^{-it \lambda/\hbar} \nonumber \\
&=&Tr_{b} \left\{ e^{-\hat S} e^{\hat S(t)}\hat p (t)  \hat \rho_{b}\right\} e^{-it \lambda/\hbar}.  \label{eq:cp1-tr}
\een
Note that $\hat p(t)$ is proportional to $\hat S(t)$ as follows:
\be
\hat p (t)=\sqrt{\frac{\hbar\omega}{2}}i \left (\hat b^\dagger e^{i\omega t} - b e^{-i\omega t} \right )=\sqrt{\frac{\hbar\omega}{2} }\frac{i}{g} \hat S(t) \label{eq:p-s-t}
\ee
Therefore, the trace operation in eq. \ref{eq:cp1-tr} can be expressed as
\ben
&&Tr_{b} \left\{ e^{-\hat S} e^{\hat S(t)}\hat p (t)  \hat \rho_{b}\right\} \nonumber \\
&&=\sqrt{\frac{\hbar\omega}{2}} \frac{i}{g}\frac{\partial}{\partial \alpha} Tr_b \left \{ e^{-\hat S}e^{\alpha \hat S (t)} \hat \rho_b\right \}\Big |_{\alpha=1} .
\een
Following a procedure similar to obtaining eq. \ref{eq:sst-trace}, we can calculate the trace in the above expression as follows:
\ben
&&Tr_b \left \{ e^{-\hat S}e^{\alpha \hat S (t)} \hat \rho_b\right \} \nonumber \\
&&=Tr_{b} \left\{ e^{-\hat S+\alpha\hat S(t)} \hat \rho_{b}\right\}e^{-\alpha[\hat S,\hat S(t)]/2} \nonumber \\
&&=Tr_{b} \left\{ e^{g\hat b (1-\alpha e^{-i\omega t})-g\hat b^\dagger (1-\alpha e^{i\omega t} )} \hat \rho_{b}\right\}e^{i\alpha g^2\sin (\omega t)} \nonumber \\
&&=Tr_{b} \left\{ e^{g\hat b (1-\alpha e^{-i\omega t})}e^{-g\hat b^\dagger (1-\alpha e^{i\omega t} )} \hat \rho_{b}\right\}\nonumber \\
&&\hspace{1in}\times e^{g^2|(1-\alpha e^{-i\omega t}) |^2/2+i\alpha g^2\sin (\omega t)} \nonumber \\
&&=e^{-g^2|1-\alpha e^{-i\omega t}|^2/(1- e^{-\beta\hbar\omega })} e^{g^2|1-\alpha e^{-i\omega t} |^2/2}e^{i\alpha g^2\sin (\omega t)} \nonumber \\
&&=e^{-g^2[ \coth(\beta\hbar\omega/2) \{(1+\alpha^2)-2\alpha \cos (\omega t)\}/2-i\alpha \sin (\omega t)]}  . \label{eq:tr-s-ast}
\een
Therefore, 
\ben
&&C_p^{(1)}(t)=e^{-i\lambda t/\hbar} \sqrt{\frac{\hbar\omega}{2}} \frac{i}{g}\nonumber \\
&&\times \frac{\partial}{\partial \alpha} e^{-g^2[ \coth(\beta\hbar\omega/2) \{(1+\alpha^2)-2\alpha \cos (\omega t)\}/2-i\alpha \sin (\omega t)]}\Big |_{\alpha=1} \nonumber \\
&&=\sqrt{\frac{\hbar\omega}{2}}\frac{1}{ig} {\mathcal K}_g(t) e^{-{\mathcal K}_g(t)-i\lambda t/\hbar} . \label{eq:cp1}
\een

Second, let us consider the following time correlation function:
\be
C_p^{(2)}(t)=Tr_{b} \left\{ \hat p e^{it (\hat H_{b}+\hat B)/\hbar} e^{-it \hat H_{b}/\hbar} \hat \rho_{b}\right\} .
\ee
Following a procedure similar to eq. \ref{eq:trbj-1}, we find that 
\ben
C_p^{(2)}(t)&=&Tr_{b} \left\{ \hat p e^{-\hat S} e^{\hat S}e^{it (\hat H_{b}+\hat B)/\hbar} e^{-\hat S} e^{\hat S} e^{-it \hat H_{b}/\hbar} \hat \rho_{b}\right\} \nonumber \\
&=&Tr_{b} \left\{ \hat p e^{-\hat S} e^{it \hat H_{b}/\hbar} e^{\hat S} e^{-it \hat H_{b}/\hbar} \hat \rho_{b}\right\} e^{-it \lambda/\hbar} \nonumber \\
&=&Tr_{b} \left\{ \hat p e^{-\hat S} e^{\hat S(t)} \hat \rho_{b}\right\} e^{-it \lambda/\hbar}.  \label{eq:cp2-tr}
\een
Employing Eq. (\ref{eq:p-s-t}) for $t=0$, we find that the trace in the last line of the above equation can be expressed as
\ben
&&Tr_{b} \left\{ \hat p e^{-\hat S} e^{\hat S(t)} \hat \rho_{b}\right\} \nonumber \\
&&=-\sqrt{\frac{\hbar\omega}{2}} \frac{i}{g}\frac{\partial}{\partial \alpha} Tr_b \left \{ e^{-\alpha \hat S}e^{\hat S (t)} \hat \rho_b\right \}\Big |_{\alpha=1} .
\een
The trace in the above expression can be calculated in a way similar to eq. \ref{eq:tr-s-ast} as follows:
\ben
&&Tr_b \left \{ e^{-\alpha \hat S}e^{\hat S (t)} \hat \rho_b\right \} \nonumber \\
&&=Tr_{b} \left\{ e^{-\alpha \hat S+\hat S(t)} \hat \rho_{b}\right\}e^{-\alpha[\hat S,\hat S(t)]/2} \nonumber \\
&&=Tr_{b} \left\{ e^{g\hat b (\alpha-e^{-i\omega t})-g\hat b^\dagger (\alpha- e^{i\omega t} )} \hat \rho_{b}\right\}e^{i\alpha g^2\sin (\omega t)} \nonumber \\
&&=Tr_{b} \left\{ e^{g\hat b (\alpha- e^{-i\omega t})}e^{-g\hat b^\dagger (\alpha- e^{i\omega t} )} \hat \rho_{b}\right\}\nonumber \\
&&\hspace{1in}\times e^{g^2|(\alpha- e^{-i\omega t}) |^2/2+i\alpha g^2\sin (\omega t)} \nonumber \\
&&=e^{-g^2|\alpha- e^{-i\omega t}|^2/(1- e^{-\beta\hbar\omega })} e^{g^2|\alpha- e^{-i\omega t} |^2/2}e^{i\alpha g^2\sin (\omega t)} \nonumber \\
&&=e^{-g^2[ \coth(\beta\hbar\omega/2) \{(\alpha^2+1)-2 \alpha \cos (\omega t)\}/2-i\alpha \sin (\omega t)]} .\label{eq:tr-as-st}
\een
Therefore, 
\ben
&&C_p^{(2)}(t)=-e^{-i\lambda t/\hbar} \sqrt{\frac{\hbar\omega}{2}} \frac{i}{g}\nonumber \\
&&\times \frac{\partial}{\partial \alpha} e^{-g^2[ \coth(\beta\hbar\omega/2) \{(\alpha^2+1)-2\alpha \cos (\omega t)\}/2-i\alpha \sin (\omega t)]}\Big |_{\alpha=1} \nonumber \\
&&=\sqrt{\frac{\hbar\omega}{2}}\frac{i}{g} {\mathcal K}_g(t) e^{-{\mathcal K}_g(t)-i\lambda t/\hbar} . \label{eq:cp2}
\een

Finally, let us consider the following momentum correlation function:
\ben
C_{pp}(t)&=&Tr_{b} \left\{ e^{it (\hat H_{b}+\hat B)/\hbar} \hat p e^{-it \hat H_{b}/\hbar} \hat \rho_{b}\hat p\right\} \nonumber \\
&=&Tr_{b} \left\{ \hat p e^{it (\hat H_{b}+\hat B)/\hbar} \hat p e^{-it \hat H_{b}/\hbar} \hat \rho_{b}\right\} .
\een
Following a procedure similar to eq. \ref{eq:trbj-1}, we find that
\ben
C_{pp}(t)&=&Tr_{b} \left\{ \hat p e^{-\hat S} e^{\hat S}e^{it (\hat H_{b}+\hat B)/\hbar} e^{-\hat S} e^{\hat S} \hat p e^{-it \hat H_{b}/\hbar} \hat \rho_{b}\right\} \nonumber \\
&=&Tr_{b} \left\{ \hat p e^{-\hat S} e^{it \hat H_{b}/\hbar} e^{\hat S} \hat p e^{-it \hat H_{b}/\hbar} \hat \rho_{b}\right\} e^{-it \lambda/\hbar} \nonumber \\
&=&Tr_{b} \left\{ \hat p e^{-\hat S} e^{\hat S(t)} \hat p(t)\hat \rho_{b}\right\} e^{-it \lambda/\hbar} .  \label{eq:cp2-tr}
\een
In the above expression, the trace can be expressed as 
\ben
&&Tr_{b} \left\{ \hat p e^{-\hat S} e^{\hat S(t)} \hat p(t)\hat \rho_{b}\right\} \nonumber \\
&&=\frac{\hbar\omega}{2g^2}\frac{\partial^2}{\partial \alpha_1\partial \alpha_2} \left . Tr_{b} \left\{ e^{-\alpha_1 \hat S} e^{\alpha_2 \hat S(t)}\hat \rho_{b}\right\}\right|_{\alpha_1=\alpha_2=1}  . \label{eq:ppt-gen}
\een
Following a procedure similar to eqs. \ref{eq:tr-s-ast} and \ref{eq:tr-as-st}, the trace in the above expression can be calculated as follows:
\ben
&&Tr_{b} \left\{ e^{-\alpha_1 \hat S} e^{\alpha_2 \hat S(t)}\hat \rho_{b}\right\} \nonumber \\
&&=Tr_{b} \left\{ e^{-\alpha_1 \hat S+\alpha_2\hat S(t)} \hat \rho_{b}\right\}e^{-\alpha_1\alpha_2[\hat S,\hat S(t)]/2} \nonumber \\
&&=Tr_{b} \left\{ e^{-g\hat b (\alpha_1-\alpha_2 e^{-i\omega t})+g\hat b^\dagger (\alpha_1 - \alpha_2 e^{i\omega t} )} \hat \rho_{b}\right\}e^{i\alpha_1\alpha_2 g^2\sin (\omega t)} \nonumber \\
&&=Tr_{b} \left\{ e^{-g\hat b (\alpha_1-\alpha_2 e^{-i\omega t})}e^{g\hat b^\dagger (\alpha_1- \alpha_2 e^{i\omega t} )} \hat \rho_{b}\right\}\nonumber \\
&&\hspace{1in}\times e^{g^2|(\alpha_1- \alpha_2 e^{-i\omega t}) |^2/2+i\alpha_1\alpha_2 g^2\sin (\omega t)} \nonumber \\
&&=e^{-g^2|\alpha_1- \alpha_2e^{-i\omega t}|^2/(1- e^{-\beta\hbar\omega })} e^{g^2|\alpha_1- \alpha_2 e^{-i\omega t} |^2/2}e^{i\alpha_1\alpha_2 g^2\sin (\omega t)} \nonumber \\
&&=e^{-g^2[ \coth(\beta\hbar\omega/2) \{(\alpha_1^2+\alpha_2^2-2\alpha_1\alpha_2 \cos (\omega t)\}/2-i\alpha_1\alpha_2 \sin (\omega t)]} \nonumber \\
\een
Taking partial derivatives of the above expression with respect to $\alpha_1$ and $\alpha_2$, we find that
\ben
&&\left . \frac{\partial^2}{\partial \alpha_1\partial \alpha_2}Tr_{b} \left\{ e^{-\alpha_1 \hat S} e^{\alpha_2 \hat S(t)} \hat \rho_{b}\right\} \right|_{\alpha_1=\alpha_2=1} \nonumber \\
&&=\left \{ g^2 \left (\coth(\frac{\beta\hbar\omega}{2})\cos (\omega t)+i\sin (\omega t) \right)+K_g(t)^2 \right\} e^{-K_g(t)} ,\nonumber \\
\een
where $K_g(t)$ is defined by eq. \ref{eq:kgt}.
Employing the above expression in eq. \ref{eq:ppt-gen} and then using eq. \ref{eq:cp2-tr}, we obtain the following expression:
\ben
C_{pp}(t)&=&\frac{\hbar\omega}{2g^2}\Big \{ g^2 \left (\coth(\frac{\beta\hbar\omega}{2})\cos (\omega t)+i\sin (\omega t) \right)\nonumber \\
&&+K_g(t)^2 \Big \}e^{-K_g(t)-it\lambda/\hbar} . \label{eq:cpp}
\een

\subsection{Evaluation of eq. \ref{eq:cjj't-1}}
\re{This section describes how the momentum bath correlation function, eq. \ref{eq:cjj't-1}, can be evaluated.
For the case where $j\neq j'$, it can be calculated as follows:
\ben
{\mathcal C}_{jj'}(t)&=&{\rm Tr}_{bj} \left \{e^{i t(\hat H_{bj}+\hat B_j)/\hbar} \hat p_j e^{-i t\hat H_{bj}/\hbar} \hat \rho_{bj}\right\} \nonumber \\
&&\times {\rm Tr}_{bj'} \left \{e^{i t(\hat H_{bj'}+\hat B_{j'})/\hbar} e^{-i t\hat H_{bj'}/\hbar} \hat \rho_{bj'} \hat p_{j'} \right\}  \nonumber \\ 
&&\times \prod_{j'' \neq j,j'} {\rm Tr}_{bj''} \left \{e^{i t(\hat H_{bj''}+\hat B_{j''})/\hbar} e^{-i t\hat H_{bj''}/\hbar} \hat \rho_{bj''} \right\}  \nonumber \\
&&=\frac{\hbar\sqrt{\omega_j\omega_{j'}}}{2g_jg_{j'}}{\mathcal K}_j(t){\mathcal K}_{j'}(t) e^{-{\mathcal K}(t) -i\lambda t/\hbar} ,
\een
where eqs. \ref{eq:ct}, \ref{eq:cp1}, or \ref{eq:cp2} in Appendix A have been used for each relevant vibrational mode, and \ret{the reorganization energy $\lambda$ and} bath correlation functions are defined as
\ben
&&\lambda=\sum_j \hbar\omega_j g_j^2 ,\\
&&{\mathcal K}_j(t)= g_j^2 \left (\coth(\frac{\beta\hbar\omega_j}{2}) (1-\cos(\omega_j t)) -i\sin(\omega_{j} t )\right ) , \nonumber  \\ \\
&&{\mathcal K}(t)=\sum_j {\mathcal K}_j(t)  . \label{eq:kt-def}
\een 
For the case where $j=j'$, eq. \ref{eq:cjj't-1} can be shown to be 
\ben
{\mathcal C}_{jj}(t)&=&{\rm Tr}_{bj} \left \{e^{i t(\hat H_{bj}+\hat B_j)/\hbar} \hat p_j e^{-i t\hat H_{bj}/\hbar} \hat \rho_{bj} \hat p_j\right\} \nonumber \\
&&\times \prod_{j' \neq j} {\rm Tr}_{bj'} \left \{e^{i t(\hat H_{bj'}+\hat B_{j'})/\hbar} e^{-i t\hat H_{bj'}/\hbar} \hat \rho_{bj'} \right\} \nonumber \\
&=&\frac{\hbar\omega_j}{2}\Bigg \{\frac{{\mathcal K}_j(t)^2}{g_j^2} \nonumber \\
&&+\left (\coth(\frac{\beta\hbar\omega_j}{2})\cos (\omega_jt)+i\sin(\omega_j t)\right) \Bigg \} \nonumber \\
&&\times e^{-{\mathcal K}(t) -i\lambda t/\hbar} ,
\een
where eq.  \ref{eq:ct} or \ref{eq:cpp} has been used.  }

\re{Employing the above expressions in eq. \ref{eq:cjj't-1}, the FGR rate can be expressed in the form of eq. \ref{eq:fgr-2} with the following expressions for the additional bath correlation functions: 
\ben
F(t)&=&\sum_j \left (\frac{\hbar \omega_j}{2}\right)^{1/2} \tilde F_{j,12} g_j \Big (\coth(\frac{\beta\hbar\omega_j}{2}) (1-\cos (\omega_j t))\nonumber \\
&&\hspace{1in}-i\sin (\omega_j t) \Big ) , \label{eq:ft-def}
\een
$F^*(t)$, which  is complex conjugate of $F(t)$, and
\ben
D(t)&=&\sum_j \frac{\hbar\omega_j}{2} | \tilde F_{j,12}|^2 \nonumber \\
&&\times \Big (\coth(\frac{\beta\hbar\omega_j}{2}) \cos (\omega_j t))+i\sin (\omega_j t) \Big ) .  \label{eq:dt-def}
\een
}

\ \vspace{5in}\\

\newpage
\newpage

\renewcommand{\theequation}{S\arabic{equation}}
\renewcommand{\thefigure}{S\arabic{figure}}
\renewcommand{\thetable}{S\arabic{table}}
\renewcommand\thepage{S\arabic{page}}
\setcounter{page}{1}
\setcounter{figure}{0}
\setcounter{equation}{0}

\newpage

\begin{widetext}
\begin{center}
{\bf \Large Supporting Information: Fermi's golden rule rate expression for transitions due to nonadiabatic derivative couplings in the adiabatic basis}
\end{center}
\end{widetext}






\noindent
{\bf SI. Hamiltonians in the adiabatic basis and nonadiabatic derivative couplings} \vspace{.2in}\\
{\bf 1. Isolated molecules} \vspace{.2in}\\
{\bf a. Hamiltonian} \vspace{.2in}\\
Consider a molecular Hamiltonian with $N_e$ electrons and $N_u$ nuclei.  For now, let us assume\footnote{This assumption will be relaxed later once adiabatic basis of molecular states are identified.} that there are no other degrees of freedom involved.   The corresponding Hamiltonian in atomic units can be expressed as\cite{jang-qmc} 
\be
\hat H=\hat H_{en}+\hat H_{nu} , \label{eq:h-mol}
\ee
where 
\ben
&&\hat H_{en}=\sum_{\mu=1}^{N_e} \frac{\hat {\bf p}_\mu^2}{2} -\sum_{\mu=1}^{N_e}\sum_{c=1}^{N_u} \frac{Z_c e^2}{|\hat {\bf r}_\mu-\hat {\bf R}_c|^2} +\frac{1}{2}\sum_{\mu=1}^{N_e}\sum_{\nu\neq \mu} \frac{1}{|\hat {\bf r}_\mu-\hat {\bf r}_\nu|} ,\nonumber \\  \label{eq:h-en}\\
&&\hat H_{nu}=\sum_{c=1}^{N_u} \frac{\hat {\bf P}_c}{2M_c}+\frac{1}{2}\sum_{c=1}^{N_u}\sum_{c'\neq c} \frac{Z_cZ_{c'}}{|\hat {\bf R}_c-\hat {\bf R}_{c'}|}  .
\een
In the above expressions, $\hat {\bf p}_\mu$ and $\hat {\bf r}_\mu$ are momentum and position operators of an electron labeled with $\mu$, and $\hat {\bf P}_c$, $\hat {\bf R}_c$,  $M_c$, and $Z_c$ are momentum operator, position operator, mass, and charge of a nucleus labeled with $c$. 
 
Let us consider the adiabatic electronic Hamiltonian $\hat H_{en}({\bf R})$, which is the same as  Eq. (\ref{eq:h-en}) except that the nuclear position operator vector $\hat {\bf R}\equiv (\hat {\bf R}_1,\cdots, \hat {\bf R}_{N_u})^T$ in Eq. (\ref{eq:h-en}) is replaced with corresponding vector parameter: ${\bf R}\equiv ( {\bf R}_1,\cdots, {\bf R}_{N_u})^T$.  Then, one can define the following adiabatic electronic states and eigenvalues:
\be
\hat H_{en}({\bf R})|\psi_{e,k}({\bf R})\rangle =E_{e,k}({\bf R})|\psi_{e,k}({\bf R})\rangle ,
\ee
where $k$ collectively denotes the set of all quantum numbers that are necessary to completely specify the adiabatic electronic states defined for the given nuclear coordinate ${\bf R}$.  Thus, given that all such states can be identified, the following completeness relation holds in the electronic space:
\be
\hat 1_{e,{\bf R}}=\sum_k |\psi_{e,k}({\bf R})\rangle \langle \psi_{e,k}({\bf R})| . \label{eq:elec-id-res}
\ee
In the above expression, the subscript ${\bf R}$ denotes that the resolution is with respect to adiabatic electronic states defined at ${\bf R}$.  This subscript is used to make it clear that each component of the resolution is dependent on ${\bf R}$ even though in principle the electronic identity operator itself  is independent of nuclear coordinates.  In practice,  due to the approximations made for the adiabatic electronic states and the truncation of the summation, the expansion is not complete.    Thus, in practice, the righthand side of Eq. (\ref{eq:elec-id-res}) ends up  being an approximation and becomes dependent upon ${\bf R}$.

At a formal level, it is always possible to assume the existence of Eq. (\ref{eq:elec-id-res}).   When this identity resolution is used carefully, it is possible\cite{jang-jcp-nonad} to decompose the molecular Hamiltonian $\hat H$ into adiabatic and nonadiabatic components such that both are expressed in terms of outer products (in Dirac notation) involving adiabatic electronic states and nuclear position states in a consistent manner.  Thus, the molecular Hamiltonian, Eq. (\ref{eq:h-mol}), can be expressed as\footnote{The original notations were slightly altered and $\hbar$ was incorporated into derivative coupling terms  in this work.} 
\be
\hat H=\hat H_{ad}+\frac{1}{2} \sum_{\alpha=1}^{3N_u} \left ( \hat P_\alpha \hat F_\alpha+\hat F_\alpha \hat P_\alpha \right)+ \hat S , \label{eq:mol-ad}
\ee
where $\alpha$ denotes each one dimensional Cartesian component and $\hat H_{ad}$ is the adiabatic approximation for the molecular Hamiltonian given by
\ben
\hat H_{ad}&=&\int d{\bf R} \sum_k |{\bf R}\rangle |\psi_{e,k} ({\bf R})\rangle \Big \{ -\sum_{\alpha=1}^{3N_u} \frac{\hbar^2}{2M_\alpha} \frac{\partial^2}{\partial R_\alpha^2}  \nonumber \\ &&\hspace{1in}+U_k({\bf R})\Big \}\langle \psi_{e,k} ({\bf R})|\langle {\bf R}| ,
\een
with
\be
U_k ({\bf R})=E_{e,k} ({\bf R}) +\frac{1}{2}\sum_{c=1}^{N_u}\sum_{c'\neq c} \frac{Z_cZ_{c'}}{|{\bf R}_c- {\bf R}_{c'}|} . 
\ee 
In Eq. (\ref{eq:mol-ad}),  $\hat F_\alpha$ represents the first nonadiabatic derivative coupling (NDC) term,  
\be
\hat F_\alpha=\int d{\bf R} \sum_k\sum_{k'} |{\bf R}\rangle |\psi_{e,k}({\bf R})\rangle F_{\alpha,kk'}({\bf R})\langle \psi_{e,k'} ({\bf R})| \langle {\bf R}| , \label{eq:f-alpha}
\ee
where
\be
F_{\alpha,kk'} ({\bf R})=\frac{\hbar}{i M_\alpha} \langle \psi_{e,k}({\bf R})| \left (\frac{\partial}{\partial R_\alpha} |\psi_{e,k'}({\bf R})\rangle \right) . \label{eq:f-alpha-kkp-0}
\ee
Note that $\hat P_\alpha\hat F_\alpha$ in Eq. (\ref{eq:mol-ad}) contains part of the conventional second NDC term.  
The last term in Eq. (\ref{eq:mol-ad}) represents  the remaining second NDC term,
\be
\hat S=\int d{\bf R} \sum_k\sum_{k'} |{\bf R}\rangle |\psi_{e,k}({\bf R})\rangle S_{kk'}({\bf R})\langle \psi_{e,k'} ({\bf R})| \langle {\bf R}| , \label{eq:s-op}
\ee
where each component involves products of  first NDC terms\cite{jang-jcp-nonad} as follows:
\be
S_{kk'}({\bf R})=\frac{1}{2}\sum_{\alpha=1}^{3N_u} \sum_{k''} M_\alpha F_{\alpha,kk''}({\bf R}) F_{\alpha,k''k'}({\bf R}) .
\ee
Thus, Eq. (\ref{eq:mol-ad}) with above clarification shows that all NDC terms can be determined once full information on functional form of Eq. (\ref{eq:f-alpha-kkp-0}) is known.  
From the condition that $\langle \psi_{e,k}({\bf R})|\psi_{e,k'}({\bf R})\rangle=\delta_{kk'}$, it is easy to show\cite{jang-jcp-nonad} that 
\be
F_{\alpha,kk'}({\bf R})=F_{\alpha,k'k}^*({\bf R}) . \label{eq:f-alpha-kkp-cc} 
\ee
Thus, it is clear that $\hat F_\alpha$ defined by Eq. (\ref{eq:f-alpha}) is Hermitian.
This also means that the diagonal component $F_{\alpha,kk}({\bf R})$ is always real.  Thus, if all the electronic eigenfunctions $\langle {\bf r}|\psi_{e,j}({\bf R})\rangle$ can be expressed as real-valued functions, $F_{\alpha,kk}({\bf R})$ defined as Eq. (\ref{eq:f-alpha-kkp-0}) is always zero.  On the other hand,  in the presence of magnetic fields or for other cases where the adiabatic electronic eigenstate needs to be complex valued,  it does not have to be zero.

The expressions given by Eqs. (\ref{eq:mol-ad})-(\ref{eq:f-alpha-kkp-0}) are equivalent to standard expressions, which are typically expressed in terms of those projected onto particular adiabatic electronic and nuclear wave functions. Compared to these expressions, our expressions here provide complete nonadiabatic coupling Hamiltonian terms as standalone Hermitian operators in the most general way, and also makes it possible to understand approximations involved dynamics employing the Hamiltonian.  \vspace{.2in} \\

\noindent
{\bf b. Eigenstates of the adiabatic Hamiltonian and their nonadiabatic couplings} \vspace{.2in} \\
Let us denote the eigenstate of the nuclear degrees of freedom for an adiabatic potential energy surface $U_k({\bf R})$ with eigenvalue $E_{n_k}$ as $|\chi_{n_k}\rangle$.  
Then, 
\be
\left (\sum_{\alpha=1}^{3N_u} \frac{\hat P_\alpha^2}{2M_\alpha} +U_k(\hat {\bf R})\right) |\chi_{n_k}\rangle=E_{n_k} |\chi_{n_k}\rangle , 
\ee
or equivalently, 
\be
\left (-\sum_{\alpha=1}^{3N_u} \frac{\hbar^2}{2M_\alpha}\frac{\partial^2}{\partial R_\alpha^2} +U_k({\bf R})\right) \chi_{n_k}({\bf R})=E_{n_k} \chi_{n_k} ({\bf R}) , 
\ee
where $\chi_{n_k}({\bf R})=\langle {\bf R}| \chi_{n_k}\rangle$. 
  Then, let us introduce the following adiabatic vibronic state defined in the direct product space of electronic and nuclear degrees of freedom: 
\ben
|\Psi_{k,n}\rangle &=&\int d{\bf R}  (|{\bf R}\rangle\langle {\bf R} |\chi_{n_k}\rangle) \otimes|\psi_{e,k}( {\bf R})\rangle \nonumber \\
&=&\int d{\bf R} \     \chi_{n_k}({\bf R}) |{\bf R}\rangle |\psi_{e,k}( {\bf R})\rangle , 
\een
where in the second line we have omitted the direct product symbol $\otimes$ between   $|{\bf R}\rangle$ and  $|\psi_{e,k}( {\bf R})\rangle$. 
Then,  application of $\hat H_{ad,k}$ to $|\Psi_{k,n}\rangle$ results in
\ben
&&\hat H_{ad,k} |\Psi_{k,n}\rangle\nonumber \\
&=&\int d{\bf R} \int d{\bf R}' |{\bf R}\rangle |\psi_{e,k}({\bf R})\rangle   \left \{-\sum_{\alpha=1}^{3N_u} \frac{\hbar^2}{2M_\alpha} \frac{\partial^2}{\partial R_\alpha^2}+U_k({\bf R})\right\} \nonumber \\
 &&\hspace{.5in}\times \langle \psi_{e,k}({\bf R}) |\psi_{e,k}({\bf R}')\rangle \langle {\bf R}  |{\bf R}'\rangle \chi_{n_k}({\bf R}')\nonumber \\
 &=&\int d{\bf R} |{\bf R}\rangle |\psi_{e,k}({\bf R})\rangle   \left \{-\sum_{\alpha=1}^{3N_u} \frac{\hbar^2}{2M_\alpha} \frac{\partial^2}{\partial R_\alpha^2}+U_k({\bf R})\right\} \chi_{n_k}({\bf R})\nonumber \\
  &=&E_{n_k}\int d{\bf R} |{\bf R}\rangle |\psi_{e,k}({\bf R})\rangle  \chi_{n_k}({\bf R}) 
  =E_{n_k}  |\Psi_{k,n}\rangle .
 \een 
 Thus, $|\Psi_{k,n}\rangle$ is an eigenstate of $\hat H_{ad,k}$.  In a similar manner, it is straightforward to show that $\hat H_{ad,k'} |\Psi_{k,n}\rangle =0$ for $k\neq k'$. 
 
 Now let us consider application of nonadiabatic terms as follows.  
 \begin{widetext}
 \ben
 &&\frac{1}{2}\sum_{\alpha=1}^{3N_u} \left (\hat P_\alpha \hat F_\alpha+\hat F_\alpha\hat P_\alpha \right ) |\Psi_{k,n}\rangle =\frac{1}{2}\sum_{\alpha=1}^{3N_u} \sum_{k'}\sum_{k''}\int d{\bf R} \int d{\bf R}' \left \{ \hat P_\alpha  |{\bf R}'\rangle |\psi_{e,k'}({\bf R}')\rangle F_{\alpha,k'k''}({\bf R}')\langle \psi_{e,k''} ({\bf R}')| \langle {\bf R}'| \right . \nonumber \\
 &&\left .\hspace{1 in} +  |{\bf R}'\rangle |\psi_{e,k'}({\bf R}')\rangle F_{\alpha,k'k''}({\bf R}')\langle \psi_{e,k''} ({\bf R}')| \langle {\bf R}'| \hat P_\alpha \right\}  |\psi_{e,k}( {\bf R})\rangle |{\bf R}\rangle \langle {\bf R}|\chi_{n_k}\rangle \nonumber \\
  &&\hspace{.5in}=\frac{1}{2}\sum_{\alpha=1}^{3N_u} \Bigg \{ \sum_{k'}\int d{\bf R}  \hat P_\alpha  |{\bf R}\rangle |\psi_{e,k'}({\bf R})\rangle F_{\alpha,k'k}({\bf R}) \langle {\bf R}| \chi_{n_k}\rangle \nonumber \\
 &&\hspace{1 in} + \sum_{k'}\sum_{k''}\int d{\bf R}\int d{\bf R}' |{\bf R}'\rangle |\psi_{e,k'}({\bf R}')\rangle F_{\alpha,k'k''}({\bf R}')\langle \psi_{e,k''} ({\bf R}')|  \psi_{e,k}( {\bf R})\rangle\langle {\bf R}'| \hat P_\alpha |{\bf R}\rangle \langle {\bf R}  |\chi_{n_k}\rangle \Bigg \} .
 \een 
 Taking inner product of the above equation with  $\langle \Psi_{l,m}|=\int d{\bf R}'' \   \langle \psi_{e,l}( {\bf R}'')|\langle {\bf R}''| \chi_{l_m}^*({\bf R}'')=\int d{\bf R}'' \   \langle \psi_{e,l}( {\bf R}'')|\langle \chi_{l_m}|{\bf R}''\rangle\langle {\bf R}''|$ on the right hand side, we obtain
  \ben
 &&\langle \Psi_{l,m}| \frac{1}{2}\sum_{\alpha=1}^{3N_u} \left (\hat P_\alpha \hat F_\alpha+\hat F_\alpha\hat P_\alpha \right )  |\Psi_{k,n}\rangle =\frac{1}{2}\sum_{\alpha=1}^{3N_u} \Bigg \{ \sum_{k'}\int d{\bf R} \int d{\bf R}''\langle \chi_{l_m}|{\bf R}''\rangle \langle {\bf R}''| \hat P_\alpha  |{\bf R}\rangle\langle \psi_{e,l} ({\bf R}'') |\psi_{e,k'}({\bf R})\rangle F_{\alpha,k'k}({\bf R}) \langle {\bf R}| \chi_{n_k}\rangle \nonumber \\
 &&\hspace{1 in} + \sum_{k''}\int d{\bf R}\int d{\bf R}''  \langle \chi_{l_m}|{\bf R}''\rangle F_{\alpha,lk''}({\bf R}'')\langle \psi_{e,k''} ({\bf R}'')|  \psi_{e,k}( {\bf R})\rangle\langle {\bf R}''| \hat P_\alpha |{\bf R}\rangle \langle {\bf R}  |\chi_{n_k}\rangle \Bigg \} .
 \een 
 Noting that $\langle {\bf R}''|\hat P_\alpha|{\bf R}\rangle=(\hbar/i) (\partial/ \partial R_\alpha'') \delta ({\bf R}''-{\bf R})=-(\hbar/i) (\partial/ \partial R_\alpha) \delta ({\bf R}''-{\bf R}) $ and then conducting integration by part, we obtain
   \ben
 &&\langle \Psi_{l,m}| \frac{1}{2}\sum_{\alpha=1}^{3N_u} \left (\hat P_\alpha \hat F_\alpha+\hat F_\alpha\hat P_\alpha \right )  |\Psi_{k,n}\rangle \nonumber \\
 &&=\frac{\hbar}{2i}\sum_{\alpha=1}^{3N_u} \Bigg \{ \sum_{k'}\int d{\bf R} \int d{\bf R}''\langle \chi_{l_m}|{\bf R}''\rangle \delta( {\bf R}''-{\bf R})\frac{\partial}{\partial R_\alpha} \langle \psi_{e,l} ({\bf R}'') |\psi_{e,k'}({\bf R})\rangle F_{\alpha,k'k}({\bf R}) \langle {\bf R}| \chi_{n_k}\rangle \nonumber \\
 &&\hspace{0.5 in} + \sum_{k''}\int d{\bf R}\int d{\bf R}''  \langle \chi_{l_m}|{\bf R}''\rangle F_{\alpha,lk''}({\bf R}'')  \delta( {\bf R}''-{\bf R})\frac{\partial}{\partial R_\alpha}\langle \psi_{e,k''} ({\bf R}'')|  \psi_{e,k}( {\bf R})\rangle\langle {\bf R}  |\chi_{n_k}\rangle \Bigg \} \nonumber \\
  &&=\frac{\hbar}{2i}\sum_{\alpha=1}^{3N_u} \Bigg \{ \int d{\bf R} \langle \chi_{l_m}|{\bf R}\rangle\left  ( \sum_{k'} \frac{iM_\alpha}{\hbar} F_{\alpha,lk'}({\bf R})  F_{\alpha,k'k}({\bf R}) \langle {\bf R}| \chi_{n_k}\rangle +\frac{\partial}{\partial R_\alpha}  F_{\alpha,lk}({\bf R})\langle {\bf R}| \chi_{n_k}\rangle \right  ) \nonumber \\
  &&\hspace{0.5 in} + \int d{\bf R}\langle \chi_{l_m}|{\bf R}\rangle \left (\sum_{k''} F_{\alpha,lk''}({\bf R}) \frac{iM_\alpha}{\hbar} F_{\alpha,k''k}({\bf R}) \langle {\bf R}  |\chi_{n_k}\rangle +F_{\alpha,lk}({\bf R})\frac{\partial}{\partial R_\alpha} \langle {\bf R}  |\chi_{n_k}\rangle \right ) \Bigg \} \nonumber \\
 &&=\sum_{\alpha=1}^{3N_u} \Bigg \{ \int d{\bf R} \chi_{l_m}^*({\bf R}) \chi_{n_k}({\bf R})\left  (\sum_{k'} M_\alpha F_{\alpha,lk'}({\bf R})  F_{\alpha,k'k}({\bf R}) +\left (\frac{\hbar}{2i}\frac{\partial} {\partial R_\alpha}  F_{\alpha,lk}({\bf R})\right ) \right )  \nonumber \\
  &&\hspace{0.5 in} + \int d{\bf R} \chi_{l_m}^*({\bf R}) F_{\alpha,lk}({\bf R})\frac{\hbar}{i} \frac{\partial}{\partial R_\alpha} \chi_{n_k}({\bf R})  \Bigg \} .
 \een 
Note the extra terms containing sum of  $F_{\alpha,lk'}({\bf R})F_{\alpha,k'k}({\bf R})$ in the above expression, which appear due to non-orthogonality of adiabatic states.  Interestingly, these are the same as  the matrix element of the remaining second derivative term $\hat S$ as shown below.
 \be
 \langle \Psi_{l,m}| \hat S |\Psi_{k,n}\rangle = \frac{1}{2} \sum_{\alpha=1}^{3N_u}  \int d{\bf R} \chi_{l_m}^*({\bf R}) \chi_{n_k}({\bf R})\left  (\sum_{k'} M_\alpha F_{\alpha,lk'}({\bf R})  F_{\alpha,k'k}({\bf R}) \right ) .
 \ee
 \end{widetext}  
\ \vspace{.2in}\\  
\noindent  
{\bf c. Helllman-Feynman expression for Derivative coupling terms} \vspace{.2in}\\
For the case where $k\neq k'$ and  for non-degenerate $E_{e,k}({\bf R})$ and $E_{e,k'}({\bf R})$, the Hellmann-Feynman theorem\cite{hellmann,feynman-pr56} can be used to obtain an alternative expression for $F_{\alpha,kk'}({\bf R})$, which is obtained by taking the derivative of the following identity:
\be
\langle \psi_{e,k}({\bf R})|\hat H_e({\bf R})|\psi_{e,k'}({\bf R})\rangle =E_{e,k}({\bf R})\delta_{kk'} .  
\ee
Taking derivative of the above identity with respect to ${R}_\alpha$, we obtain 
\ben
&&\left (\frac{\partial}{\partial R_\alpha}\langle \psi_{e,k}({\bf R})|\right)\hat H_e({\bf R})|\psi_{e,k'}({\bf R})\rangle \nonumber \\
&&+\langle \psi_{e,k}({\bf R})|\left (\frac{\partial \hat H_e({\bf R}) }{\partial R_\alpha}\right) |\psi_{e,k'}({\bf R})\rangle \nonumber \\
&&+\langle \psi_{e,k}({\bf R})|\hat H_e({\bf R})\left (\frac{\partial}{\partial R_\alpha} |\psi_{e,k'}({\bf R})\rangle\right)\nonumber \\
&&=\frac{iM_\alpha}{\hbar} F_{\alpha,kk'}({\bf R}) (E_{e,k}({\bf R})-E_{e,k'}({\bf R}))\nonumber \\
&&+\langle \psi_{e,k}({\bf R})|\frac{\partial \hat H_e({\bf R}) }{\partial R_\alpha} |\psi_{e,k'}({\bf R})\rangle=\frac{\partial E_{e,k}({\bf R})}{\partial R_\alpha}\delta_{kk'} , \label{eq:der-f-alpha}
\een
where the fact that $F_{\alpha,k'k}^*({\bf R})=F_{\alpha,kk'}({\bf R})$ has been used in the second equality. 
Since we assumed that $k\neq k'$, the righthand side of the above equation is zero.   Thus, given that $E_{e,k'}({\bf R})\neq E_{e,k}({\bf R})$, Eq. (\ref{eq:der-f-alpha}) results in the following expression:
\be
F_{\alpha,kk'}({\bf R})=\frac{i\hbar}{M_\alpha} \frac{\langle \psi_{e,k}({\bf R})|(\partial \hat H_e({\bf R})/\partial R_\alpha) |\psi_{e,k'}({\bf R})\rangle}{E_{e,k}({\bf R})-E_{e,k'}({\bf R})} .\label{eq:f-alpha-kkp}
\ee 
The above Hellmann-Feynman expression makes it easy to evaluate off-diagonal derivative coupling between non-degenerate states.  Note that the above expression also clarifies that  the derivative coupling diverges between two degenerate states unless the numerator also vanishes.  Such case of divergence is known as conical intersection, for which a wealth of both theoretical and computational studies are available now.  However, in the present work, we only consider cases where $F_{\alpha,kk'}({\bf R})$ remains finite and relatively small.  These cases become important for nonradiative decay of near infrared and short wavelength infrared dye molecules, for which nonradiative decay processes exhibit the energy gap law behavior, and for dynamics in the excited state manifold in regions far from conical intersections.    \vspace{.2in}\\

\noindent
{\bf d. Normal mode representation} \vspace{.2in}\\
Let us consider the case where the electronic state is in the adiabatic state $|\psi_1({\bf R})\rangle$, for which  $U_1({\bf R})$ is the potential energy function for the nuclear degrees of freedom.  
We assume that the nuclear coordinate ${\bf R}$ is defined in the Eckart frame\cite{eckert-pr47,louck-rmp48,pickett-jacs92,dymarsky-jcp122} with respect to ${\bf R}_1^g$.  In principle, this can be identified as follows.  First, ${\bf R}'$ and the minimum energy structure for state 1, ${{\bf R}'}_1^g$, can be defined in any center-of-mass coordinate frame with respect to which the molecule is static.  Then, applying a pseudo-rotation matrix that satisfies the second Eckart condition\cite{louck-rmp48} with respect to ${{\bf R}'}_1^{g}$, a new coordinate frame can be identified.  The nuclear coordinate vectors in this rotated frame are labeled as ${\bf R}$ and ${\bf R}_1^g$.    It is known that identifying the pseudo-rotation matrix satisfying the Eckart condition is nontrivial, but there are well-established practical procedures.\cite{louck-rmp48,pickett-jacs92,dymarsky-jcp122}

We denote the mass-weighted nuclear coordinates as $\tilde {\bf R}$ such that $\tilde R_\alpha=\sqrt{M_\alpha} R_\alpha$.
Let us denote the mass-weighted nuclear coordinates that make $U_1({\bf R})$ minimum as $\tilde {\bf R}_1^g$.   Then, expanding  $U_1({\bf R})$  around $\tilde {\bf R}_1^g$ with respect to relative mass-weighted coordinates, $\tilde {\bf R}-\tilde {\bf R}_1^g$, up to the second order and diagonalizing the resulting Hessian matrix, one can determine all of normal vibrational modes and frequencies, $q_{1,j}$ and $\omega_{1,j}$ with $j=1,\cdots, N_{v,1}$, where $N_{v,1}$ is the total number of normal mode vibrations for the vibrational motion around $\tilde {\bf R}_1^g$ in the electronic state 1.  The transformation from Cartesian coordinates to these normal modes are defined as follows:
\ben
&&q_{1,j}=\sum_\alpha L_{1,j\alpha} (\tilde R_\alpha -\tilde R_{1,\alpha}^g) \nonumber \\
&&=\sum_\alpha L_{1,j\alpha}\sqrt{M_\alpha} (R_\alpha - R_{1,\alpha}^g) , j=1,\cdots, N_{v,1} .
\een

Thus, assuming that an Eckart frame that fully decouples the rotational and vibrational degrees of freedom can be found, the nuclear Hamiltonian operator for the mass-weighted coordinates on the adiabatic electronic state surface of $1$ can be expressed  as
\ben
&&\sum_\alpha \frac{\hat {\tilde P}_\alpha^2}{2}   +U_1(\hat {\bf R}) \approx \hat H_{\rm 1, tr-rot}+U_1({\bf R}_1^g)\nonumber \\
&&\hspace{.3in} +\sum_{j=1}^{N_{v,1}} \left (\frac{1}{2}\hat p_{1,j}^2+\frac{\omega_{1,j}^2}{2}\hat q_{1,j}^2 \right) + \Delta U_1 (\hat {\bf q}_1) , \label{eq:had1-ro-vib}
\een
where $\hat p_{1,j}$ is the canonical momentum operator for $\hat q_{1,j}$, $\hat H_{\rm1, tr-rot}$ represents the translation of the center-of-mass and rotational motion around ${\bf R}_1^g$, and $\Delta U_1 (\hat {\bf q})$ is the remaining anharmonic term of the potential energy function $U_1({\bf R})$, with arguments expressed in terms of normal coordinates.     

Now, let us consider the case where the transition to adiabatic electronic state $|\psi_2 ({\bf R})\rangle$ occurs, for which the potential energy is $U_2({\bf R})$.  This can also be expanded around its minimum energy nuclear coordinates denoted as ${\bf R}_2^g$.   However, in such expansion, it is important to recognize first that ${\bf R}$ is already defined in the Eckart frame with respect to ${\bf R}_1^g$, which does not necessarily satisfy the second Eckart condition\cite{louck-rmp48} for ${\bf R}_2^g$.  This has the following two consequences:  
\begin{enumerate}
\item It is not guaranteed that the nuclear Hamiltonian defined in the electronic state  $|\psi_{e,2}({\bf R})\rangle$  can be decomposed into translation-rotation and vibrational parts as in Eq. (\ref{eq:had1-ro-vib}).  
\item  Purely vibrational displacement from ${\bf R}_1^g$ may bear some rotational component around ${\bf R}_2^g$. 

\end{enumerate}
At the moment, complete resolution of the above two issues seems not possible in general.  While these issues may be mitigated by adopting curvilinear internal coordinates, whether it results in actual advantage is not clear.\cite{min-bkcs2023}  Thus, we use Cartesian coordinates and invoke additional assumptions here.  First, we assume that the non-uniqueness\cite{louck-rmp48,dymarsky-jcp122} in the choice of the Eckart frame for ${\bf R}_1^g$ can be utilized such that ${\bf R}_2^g$ is maximally aligned with ${\bf R}_1^g$.  This will minimize the coupling term between the rotation and vibration parts for the displacement around ${\bf R}_2^g$, which we assume to be small enough and can thus  be  ignored.    Similarly, we assume that the projection of pure vibrational components around ${\bf R}_1^g$ onto rotational part around ${\bf R}_2^g$ can be discarded.  

With approximations and assumptions as noted above, which can always be tested for a given molecular system,  we can expand $U_2 ({\bf R})$ with respect to $\tilde {\bf R}-\tilde {\bf R}_2^g$ around $\tilde {\bf R}_2^g$ and identify the normal mode and frequency, $q_{2,j}$ and $\omega_{2,j}$, for $j=1,\cdots, N_{v,2}$, where  $N_{v,2}$ is the total number of normal mode vibrations for the vibrational motion around ${\bf R}_2^g$.  These are related to mass-weighted cartesian coordinates in the best Eckart frame, as prescribed above, by the following transformation,
\ben
q_{2,j}&=&\sum_\alpha L_{2,j\alpha} (\tilde R_\alpha -\tilde R_{2,\alpha}^g) \nonumber \\
&=&\sum_\alpha L_{2,j\alpha}\sqrt{M_\alpha} (R_\alpha - R_{2,\alpha}^g) , j=1,\cdots, N_{v,2} . \nonumber \\
\een
Thus, we can make the following approximation:
\ben
&&\sum_\alpha \frac{\hat {\tilde P}_\alpha^2}{2}   +U_2(\hat {\bf R}) \approx \hat H_{\rm 2, {\rm tr-rot}}+U_2({\bf R}_2^g)\nonumber \\
&&\hspace{.4in} +\sum_{j=1}^{N_{v,2}} \left (\frac{1}{2}\hat p_{2,j}^2+\frac{\omega_{2,j}^2}{2}\hat q_{2,j}^2 \right)+\Delta U_2(\hat {\bf q}_2) .
\een
In the above expression, $\hat H_{\rm2, tr-rot}$ represents the translation of the center-of-mass and rotational motion around ${\bf R}_2^g$, 
and $\Delta U_2(\hat {\bf q}_2)$ is the remaining anharmonic term that is assumed to be expressed fully in terms of normal vibrational modes. 

For the case where $N_{v,1}=N_{v,2}=N_v$, $q_{1,j}$ and $q_{2,j}$ can be related by the Duschinsky rotation matrix ${\bf J}$ and a displacement vector ${\bf K}$ as follows:
 \be
 q_{2,j}=\sum_{k=1}^{N_v}J_{jk} q_{1,k}+K_j   \ .
\ee

\ \vspace{.2in}\\
{\bf e. Expressions in the subspace of two electronic states in the quasi-adiabatic approximation} \vspace{.2in}\\
On the basis of general expressions provided up to so far, we here provide detailed expressions for approximations for the Hamiltonian in the subspace of two adiabatic electronic states within the quasi-adiabatic approximation.   For convenience, we also refer to all the nuclear degrees of freedom as bath. 
First, the zeroth order Hamiltonian can be expressed as
\ben
\hat H_0 &\approx&  \Big \{ \hat H_{1,b}+U_1({\bf R}_1^g)+S_{11}(\tilde {{\bf R}}_1^g)\Big \}|\psi_{e,1} \rangle\langle \psi_{e,1}|   \nonumber \\
&+&  \Big \{ \hat H_{2,b} +U_2({\bf R}_1^g)+S_{22}(\tilde {\bf R}_1^g)\Big \}|\psi_{e,2}\rangle \langle \psi_{e,2} | , \label{eq:h-ad-2}
\een
where
\ben
&&\hat H_{1,b}=\hat H_{1,{\rm tr-rot}}+\sum_{j=1}^{N_{v,1}} \left (\frac{1}{2}\hat p_{1,j}^2+\frac{\omega_{1,j}^2}{2}\hat q_{1,j}^2 \right) + \Delta U_1 (\hat {\bf q}_1) ,\nonumber \\ \\
&&\hat H_{2,b}=\hat H_{2,{\rm tr-rot}}+\sum_{j=1}^{N_{v,2}} \left (\frac{1}{2}\hat p_{2,j}^2+\frac{\omega_{2,j}^2}{2}\hat q_{2,j}^2 \right) + \Delta U_2 (\hat {\bf q}_2) \nonumber \\
&& \hspace{.5in} + U_2({\bf R}_2^g)-U_2({\bf R}_1^g) .
\een
Similarly,  $F_{\alpha,12}({\bf R}_{1}^g)$ can be expressed as
\be
F_{\alpha,12}({\bf R}_{1}^g)=\frac{i\hbar}{\sqrt{M_\alpha}} \frac{\langle \psi_{e,1}|\left (\partial \hat H_e({\bf R})/\partial \tilde R_\alpha\right)|_{\tilde {\bf R}=\tilde {\bf R}_1^g}  |\psi_{e,2}\rangle}{E_{e,1}({\bf R}_1^g)-E_{e,2}({\bf R}_1^g)} , \label{eq:f-alpha-2}
\ee 
where note that 
\be
E_{e,1}({\bf R}_1^g)-E_{e,2}({\bf R}_1^g) =U_1 ({\bf R}_1^g)-U_2({\bf R}_1^g) .
\ee

Let us assume that the coupling Hamiltonian $\hat H_c$ is independent of translation or rotation in the body fixed frame corresponding to the minimum energy structure for the electronic state $1$, which is consistent with neglecting the translation-rotation part of the nuclear Hamiltonian in Eq.  (\ref{eq:had1-ro-vib}).  Then, $\hat H_c$ can be expressed only in terms of those involving normal vibrational modes for the state 1 as described below.  To show this, let us first consider the following matrix element of the NDC terms within the quasi-adiabatic approximation: 
\ben
&&\langle {\bf R}| \sum_\alpha  F_{\alpha,12} ({\bf R}_1^g)|\psi_{e,1}\rangle \langle  \psi_{e,2}|\hat P_\alpha|\Psi\rangle \nonumber \\
&&=\sum_\alpha  F_{\alpha,12} ({\bf R}_1^g)|\psi_{e,1}\rangle \langle  \psi_{e,2}|\langle {\bf R}|\hat P_\alpha|\Psi\rangle \nonumber \\
&&=\sum_\alpha  F_{\alpha,12} ({\bf R}_1^g)|\psi_{e,1}\rangle \langle  \psi_{e,2}|\frac{\hbar}{i} \frac{\partial}{\partial R_\alpha} \langle {\bf R}|\Psi\rangle ,
\een
where $|\Psi\rangle$ is an arbitrary state in the total Hilbert space including both electronic and nuclear degrees of freedom. 
Given that all the translation and rotational degrees of freedom are frozen,   
\ben
\frac{\partial }{\partial R_\alpha}&=&\sum_{j=1}^{N_{v,1}} \frac{\partial q_{1,j}}{\partial R_\alpha} \frac{\partial}{\partial q_{1,j}} \nonumber \\
&=&\sum_{j=1}^{N_{v,1}} L_{1,j\alpha} \sqrt{M_\alpha} \frac{\partial}{\partial q_{1,j}} .
\een
Therefore, 
\ben 
\frac{\hbar}{i} \frac{\partial}{\partial R_\alpha} \langle {\bf R}|\Psi\rangle &= &\sum_{j=1}^{N_{v,1}} L_{1,j\alpha} \sqrt{M_\alpha} \frac{\hbar}{i}\frac{\partial}{\partial q_{1,j}} \langle {\bf R}|\Psi\rangle \nonumber \\
&=&\sum_{j=1}^{N_{v,1}} L_{1,j\alpha} \sqrt{M_\alpha} \langle {\bf R}|\hat p_{1,j} |\Psi\rangle .
\een
Similarly, $F_{\alpha,12}({\bf R}_{1}^g)$ can be expressed in terms of $\tilde F_{j,12}$ as follows:
\ben
F_{\alpha,12}({\bf R}_{1}^g)= \frac{1}{\sqrt{M_\alpha}}\sum_{j'=1}^{N_{v,1}} L_{1,\alpha j'} \tilde F_{j',12} .
\een
As a result, we find that 
\ben
&&\langle {\bf R}| \sum_\alpha  F_{\alpha,12} ({\bf R}_1^g)|\psi_{e,1}\rangle \langle  \psi_{e,2}|\hat P_\alpha|\Psi\rangle \nonumber \\
&&=\langle {\bf R}|\sum_\alpha\sum_{j=1}^{N_{v,1}}\sum_{j'=1}^{N_{v,2}}L_{1,j\alpha}L_{1,j'\alpha} |\psi_{e,1}\rangle \langle  \psi_{e,2}|\hat p_{1,j} \tilde F_{j',12}|\Psi\rangle \nonumber \\
&&=\langle {\bf R}|\sum_{j=1}^{N_{v,1}}\sum_{j'=1}^{N_{v,2}} |\psi_{e,1}\rangle \langle  \psi_{e,2}|\hat p_{1,j} \tilde F_{j',12}|\Psi\rangle , 
\een
where the fact that $\sum_\alpha  L_{1,j\alpha}L_{1,j'\alpha} =\delta_{jj'}$ has been used.  The above identity holds for an arbitrary vector ${\bf R}$, which does not have any translation and rotational degree, and for any state $|\Psi\rangle$.   Therefore, the above identity amounts to the following general identity:
\ben
&& \sum_\alpha  F_{\alpha,12} ({\bf R}_1^g)|\psi_{e,1}\rangle \langle  \psi_{e,2}|\hat P_\alpha\nonumber \\
&&=\sum_{j=1}^{N_{v,1}}\sum_{j'=1}^{N_{v,2}} |\psi_{e,1}\rangle \langle  \psi_{e,2}|\hat p_{1,j} \tilde F_{j',12} .  
\een
Combining this with its Hermitian conjugate, we obtain the following expression for $\hat H_c$: 
\ben
\hat H_c&=& \sum_{j=1}^{N_{v,1}} \hat p_{1,j} \Big \{ \tilde {F}_{j,12} |\psi_{e,1}\rangle\langle \psi_{e,2}|+\tilde F_{j,12}^* |\psi_{e,2}\rangle\langle \psi_{e,1}| \Big \}  , \nonumber \\ \label{eq:hc-2s}
\een
where
\be
\tilde F_{j,12}=i \hbar\frac{\langle \psi_{e,1}|\left (\partial \hat H_e({\bf R})/\partial  q_{1,j} |_{{\bf R}={\bf R}_1^g} \right )  |\psi_{e,2}\rangle}{E_{e,1}({\bf R}_1^g)-E_{e,2}({\bf R}_1^g)} . \label{eq:tf-j}
\ee 
\ \vspace{.2in}\\

\noindent
{\bf 2. Molecules in liquid or solid environments} \vspace{.2in}\\
For molecules in liquid or solid environments, translation, rotational, and vibrational modes of molecules are in general coupled to those of environmental degrees of freedom.  
On the other hand, since the total number of degrees of freedom for the system plus environment is virtually infinite, we can ignore its translation and rotational motion.   Thus, all degrees of freedom can be viewed as vibrational.  Let us denote the position vector for the environmental degrees of freedom collectively as ${\bf X}$.   Then, following a procedure similar to obtaining the Hamiltonian for isolated molecules, one can obtain following expressions for the zeroth order Hamiltonian term:
\ben
\hat H_0&\approx&  \Big \{ \sum_{\alpha=1}^{3N_u} \frac{\hat P_\alpha^2}{2M_\alpha}   +\sum_\xi \frac{\hat P_{\xi}^2}{2m_\xi} \nonumber \\
&&\hspace{.1in}+U_1(\hat {\bf R},\hat {\bf X})+S_{11}({\bf R}_1^g,{\bf X}_1^g)\Big \}|\psi_{e,1} \rangle\langle \psi_{e,1}|   \nonumber \\
&&+  \Big \{ \sum_{\alpha=1}^{3N_u} \frac{\hat P_\alpha^2}{2M_\alpha}   +\sum_\xi \frac{\hat P_{\xi}^2}{2m_\xi} \nonumber \\
&&\hspace{.1in}+U_2(\hat {\bf R},\hat {\bf X})+S_{22}({\bf R}_1^g,{\bf X}_1^g)\Big \}|\psi_{e,2}\rangle \langle \psi_{e,2} | ,  \label{eq:h-adx-1}
\een
where $\hat P_\xi$ is the momentum conjugate to the $\xi$ component of ${\bf X}$ and $m_\xi$ is its mass.
On the other hand, the NDC terms can still be assumed to depend only on the molecular vibrational degrees of freedom directly as follows:  
\ben
\hat H_c&\approx & \sum_{\alpha=1}^{3N_u} \hat P_\alpha \left (F_{\alpha,12}({\bf R}_{1}^g, {\bf X}_1^g) |\psi_{e,1}\rangle \langle \psi_{e,2}|  \right .  \nonumber \\
&&\left . \hspace{.5in} + F_{\alpha,12}^*({\bf R}_1^g,{\bf X}_1^g)|\psi_{e,2}\rangle \langle \psi_{e,1} |   \right ) .  \label{eq:hcx-1}
\een

Making expansion of $U_k(\hat {\bf R},\hat {\bf X})$ up to the second order of displacements around ${\bf R}_k^g$ and ${\bf X}_k^g$, and determining normal modes in the extended space of molecular and environmental degrees of freedom, $\hat H_0$ and $\hat H_c$ can still be expressed in terms of these normal coordinates.  Thus, the resulting formal expression for $\hat H_0$ and $\hat H_c$ remain the same  except that detailed forms of $\Delta U_k({\bf q}_k)$s are different and each normal mode in this case is a linear combination of molecular and environmental degrees of freedom. \vspace{.2in} \\

\noindent
{\bf SII. Quadratic approximation for bath Hamiltonians} \vspace{.2in}\\
 For the case where contributions of anharmonic terms are small or negligible, quadratic approximations can be made given that subtle assumptions concerning Eckart frame are well justified.    The resulting expressions can be combined into the following expressions for the zeroth order Hamiltonian: 
\ben
\hat H_0&\approx&  \Big \{ E_{1}^0+\hat H_{b,1} \}|\psi_{e,1} \rangle\langle \psi_{e,1}|   \nonumber \\
&+&  \Big \{ E_{2}^0+\hat H_{b,2} \Big \} |\psi_{e,2}\rangle \langle \psi_{e,2} | ,  \label{eq:h0-1}
\een
where 
\ben
&&E_1^0= U_1({\bf R}_1^g)+S_{11}({\bf R}_1^g) ,\\
&&E_2^0= U_2({\bf R}_1^g)+S_{22}({\bf R}_1^g)  ,
\een
 and 
\ben
&&\hat H_{b,1}= \sum_{j=1}^{N_{v,1}}  \left (\frac{1}{2}\hat p_{1,j}^2+\frac{\omega_{1,j}^2}{2}\hat q_{1,j}^2 \right ) , \\
&&\hat H_{b,2}= \sum_{j=1}^{N_{v,2}}  \left (\frac{1}{2}\hat p_{2,j}^2+\frac{\omega_{2,j}^2}{2}\hat q_{2,j}^2 \right ) \nonumber \\
&&\hspace{.5in} +  U_2({\bf R}_2^g) - U_2({\bf R}_1^g) .
\een
\  \vspace{.2in}\\
{\bf  SIII. Brownian oscillator spectral density to represent vibrational peaks} \vspace{.2in}\\
The actual vibrational peaks become broadened due to relaxation to other modes, anharmonicity, and couplings to environments.  
Given that all of these sources of broadening can be modeled as an Ohmic bath with spectral range larger than the given vibrational frequency, the corresponding vibrational spectral density can be modeled by the following Brownian oscillator bath spectral density:\cite{garg-jcp83}
\be
J_{\rm BO}(\omega;\omega_j,\gamma)=\frac{A_j}{\gamma^2}\frac{\omega}{\left [ (\omega^2-\omega_j^2)^2/\gamma^4+4(\omega/\gamma)^2\right]} ,
\ee
where \rc{$\gamma$ is the friction coefficient of the environment\footnote{More generally, this can be assumed to be different for each vibrational mode} and }$A_j$ can be determined so that the integration of the above spectral density over $[0,\infty]$ to be normalized as follows:
\be
A_j=\left \{ \begin{array}{ccc}4\sqrt{R_j}\left [\frac{\pi}{2}+\tan^{-1} \left( \frac{(R_j-1)}{2\sqrt{R_j}} \right) \right ]^{-1}&,  &  R_j> 0  \\ 
2&, & R_j=0 \\ 
8\sqrt{-R_j}\left [ \ln \left ( \frac{1-R_j+2\sqrt{-R_j}}{1-R_j-2\sqrt{-R_j}} \right ) \right ]^{-1}&, &R_j < 0\end{array} \right . ,
\ee
with $R_j=(\omega_j/\gamma)^2-1$.  The above spectral density behaves almost like a Lorentzian near $\omega=\omega_j$ while approaching $\omega=0$ linearly and decays fast enough (\rc{as $1/\omega^3$}) for large value of $\omega$.   Thus, we use this model to account for the broadening of delta function peaks in the bath spectral densities.  For the case of ${\mathcal J}(\omega)$, the resulting expression is 
\ben
{\mathcal J}(\omega)&=&\pi\hbar \sum_j \frac{A_j}{\gamma^2} \frac{\omega\omega_j^2 g_j^2}{\left [ (\omega^2-\omega_j^2)^2/\gamma^4+4(\omega/\gamma)^2\right]} \nonumber \\
&=&2\pi^2c\hbar \sum_j \frac{A_j}{\tilde \gamma^2} \frac{\tilde \nu \tilde \nu_j^2 g_j^2}{\left [ (\tilde \nu^2-\tilde \nu_j^2)^2/{\tilde \gamma}^4+4(\tilde \nu/\tilde\gamma)^2\right]} ,\nonumber \\
\een
where in the second line $\tilde \nu=\omega/(2\pi c)$, $\tilde \nu_j=\omega_j/(2\pi c)$, and $\tilde \gamma=\gamma/(2\pi c)$.  
\rc{Although integration for the calculation of ${\mathcal K}(t)$ employing the above spectral density is well defined, care should be taken in actual numerical integration due to the fact that ${\mathcal J}(\omega)/\omega^2$ becomes singular at $\omega=0$ and because of highly oscillatory nature of the integrand in the large $\omega$ limit.  It is also important to note that the above definition of the spectral density results in small changes in the reorganization energy, which we take into consideration in the calculation of rates.}  

\rc{To handle the numerical issues noted above, we divide ${\mathcal J}(\omega)$ into two components such that 
\be
{\mathcal J}(\omega)={\mathcal J}_0(\omega) +\delta {\mathcal J}(\omega) , \label{eq:j-omega}
\ee
where ${\mathcal J}_0(\omega)$ is a combination of Ohmic bath spectral densities given by
\be
{\mathcal J}_{0}(\omega)=2A \omega e^{-\omega/\Omega_c} \left (1-\frac{1}{2} e^{-\omega/\Omega_c} \right) , 
\ee  
with 
\be
A=\pi \hbar \sum_j \frac{\gamma^2g_j^2}{\omega_j^3} A_j .   
\ee 
Since ${\mathcal J}_0(\omega)$ is the difference of two Ohmic baths with exponential cutoff, the closed form expressions for the real and imaginary parts of ${\mathcal K}(t)$ provided in the main text, Eqs. (37) and (38) with $\lambda_h=0$,  can be used for each of the two terms, resulting in the following expressions:
\ben
&&{\mathcal K}_{0,R}(t)\approx\frac{2A}{\pi \hbar} \left \{ \frac{1}{2} \ln (1+\tau_{1,0}^2 )+\ln (1+\tau_{1,1}^2) +\ln (1+\tau_{1,2}^2) \right . \nonumber \\
&&\hspace{.1in}\left . + \frac{2(1+5\theta_1/2)}{\theta_1} \left [ \tau_{1,5/2}\tan^{-1}(\tau_{1,5/2}) -\frac{1}{2} \ln (1+\tau_{1,5/2}^2) \right ]  \right \} \nonumber \\
&&\hspace{.1in}-\frac{A}{2\pi\hbar} \left \{ \frac{1}{2} \ln (1+\tau_{2,0}^2 )+\ln (1+\tau_{2,1}^2) +\ln (1+\tau_{2,2}^2) \right . \nonumber \\
&&\hspace{.1in}\left . + \frac{2(1+5\theta_1/2)}{\theta_2} \left [ \tau_{2,5/2}\tan^{-1}(\tau_{2,5/2}) -\frac{1}{2} \ln (1+\tau_{2,5/2}^2) \right ]  \right \} ,\nonumber \\ \label{eq:kr-t} \\
&&{\mathcal K}_{0,I}(t)=\frac{2A }{\pi \hbar} \tan^{-1} (\tau_{1,0})- \frac{A }{2\pi \hbar} \tan^{-1} (\tau_{1,0})  , \label{eq:ki-t-m}
\een 
with $\theta_1=\beta\hbar\Omega_c$, $\tau_{1,n}=\Omega_c t/(1+n\theta_1)$, $\theta_2=\beta\hbar\Omega_c/2$, and $\tau_{2,n}=(\Omega_c/2) t/(1+n\theta_2)$.  All the expressions provided above are valid for any choice of $\Omega_c$.  For the present work, we chose $\Omega_c=\gamma$.}
 
\rc{The leading order for the second term in Eq. (\ref{eq:j-omega}), $\delta {\mathcal J}(\omega)$, is of $\omega^3$ as follows:  
\ben
\delta {\mathcal J}(\omega)&=&{\mathcal J}(\omega)-{\mathcal J}_0(\omega) \nonumber \\
&=&\pi \hbar \sum_j A_j \frac{\gamma^2 g_j^2}{\omega_j^2} \left \{ \frac{1}{\omega_j^2} \left ( 1- \frac{2\gamma^2}{\omega_j^2}\right )+\frac{1}{\omega_c^2} \right \} \omega^3 +\cdots  \nonumber \\
\een 
Therefore, numerical integration involving $\delta {\mathcal J}(\omega)$ in the limit of $\omega\rightarrow 0+$ can be done in a simple manner using a finite interval.  As yet, care should still be taken for the limit of $\omega\rightarrow \infty$ because of strong oscillatory nature of the integrand while $\delta J(\omega)/\omega^2$ decays proportional to $1/\omega$ in the limit of  $\omega\rightarrow \infty$.  We handle this issue by linearizing only the non-oscillatory part of the integrand and then performing exact integration for each interval.   To explain this more clearly, let us consider the following integral:
\be
F_c(t;\omega_1,\omega_2)=\int_{\omega_1}^{\omega_2} d\omega f(\omega)(1-\cos(\omega t))  ,
 \ee
 where $f(\omega)$ is a smooth function that remains finite during the interval.  Then, linearizing $f(\omega)$ within the interval of $[\omega_1,\omega_2]$, the above integral can be calculated explicitly. For $t\neq 0$,  the resulting expression is 
 \ben
&& F_c(t;\omega_1,\omega_2)\approx (\omega_2-\omega_1)\frac{(f_2+f_1)}{2} \nonumber \\
 &&\hspace{.4in}-\frac{1}{t} \left ( f_2\sin(\omega_2 t)- f_1 \sin (\omega_1 t) \right) \nonumber \\
 &&\hspace{.4in} -\frac{\cos(\omega_2 t)-\cos(\omega_1 t)}{t^2}\frac{(f_2-f_1)}{(\omega_2-\omega_1)}  , \label{eq:fct-fin}
 \een
 where $f_1=f(\omega_1)$, $f_2=f(\omega_2)$.  For $t=0$,  $F_c(0;\omega_1,\omega_2)=0$. Note that the error resulting from this approximation is of order $O((\omega_2-\omega_1)^3)$.     In a similar manner, we find that, for $t\neq 0$, 
\ben
&&F_s(t;\omega_1,\omega_2)=\int_{\omega_1}^{\omega_2} d\omega f(\omega)\sin (\omega t) \nonumber \\ 
&&\hspace{.4in}\approx \frac{1}{t}\left (\cos(\omega_1 t)f_1-\cos(\omega_2 t)f_2\right ) \nonumber \\
&&\hspace{.4in}+\frac{(f_2-f_1)}{t^2} \frac{(\sin(\omega_2 t)-\sin(\omega_1 t) )}{\omega_2-\omega_1}  . \label{eq:fst-fin}
\een 
The above expression approaches the proper limit for $t=0$, $F_s(0;\omega_1,\omega_2)=0$, and also has errors of  order $O((\omega_2-\omega_1)^3)$ as well.}

 \begin{figure}
 \ \vspace{.4in}\\
\includegraphics[width=3.in]{Figures/kit.eps}
\caption{\rc{Plots of ${\mathcal K}_I(t)$ versus time for the five different cases A-E of Table 3 in the main text  for azulene, shown as black lines.  The data without bath are provided as red dashed lines for reference.}}  
\label{figures1}
\end{figure}
 
 \rc{In summary, the real and imaginary parts of the time correlation function due to $\delta {\mathcal J}(\omega)$ can be calculated as follows:
\ben
\delta {\mathcal K}_R(t)&\approx& \sum_{n=0}^{M} F_c(t;n\delta \omega, (n+1)\delta \omega) , \label{eq:kr-delta} \\
\delta {\mathcal K}_I(t)&\approx& \sum_{n=0}^{M} F_s(t;n\delta \omega, (n+1)\delta \omega) ,  \label{eq:ki-delta}
\een
where $F_c(t;n\delta \omega, (n+1)\delta \omega)$ is given by Eq. (\ref{eq:fct-fin}) with the following definition: 
\be
f(\omega) =\frac{1}{\pi\hbar}\frac{\delta {\mathcal J}(\omega)}{\omega^2} \coth(\frac{\beta\hbar \omega}{2}) .
\ee 
and  $F_s(t;n\delta \omega, (n+1)\delta \omega)$ is given by Eq. (\ref{eq:fst-fin}) with the following definition: 
\be
f(\omega) =\frac{1}{\pi\hbar}\frac{\delta {\mathcal J}(\omega)}{\omega^2}  .
\ee  }

\begin{figure}
\ \vspace{.4in}\\
\includegraphics[width=3.in]{Figures/drt.eps}
\caption{\rc{Plots of $D_{s,R}(t)= D_R(t)/\hbar^2$ (in the unit of ${\rm ps^{-2}}$ versus time  for the two different cases B and D of Table 3 in the main text for azulene, shown as black lines.  The data without bath are provided as red dashed lines for reference.}}  
\label{figures2}
\end{figure}

\begin{figure}
\ \vspace{.4in} \\
\includegraphics[width=3.in]{Figures/dit.eps}
\caption{\rc{Plots of $D_{s,I}(t) =D_I(t)/\hbar^2$ (in the unit of ${\rm ps^{-2}}$) versus time  for the two different cases B and D of Table 3 in the main text for azulene, shown as black lines.    The data without bath are provided as red dashed lines for reference.}}  
\label{figures3}
\end{figure}

\begin{figure}
\ \vspace{.4in}\\
\includegraphics[width=3.in]{Figures/frt.eps}
\caption{\rc{Plots of $F_{s,R}(t) =F_R(t)/\hbar$ (in the unit of ${\rm ps^{-1}}$) versus time  for the two different cases B and D of Table 3 in the main text for azulene, shown as black lines.  The data without bath are provided as red dashed lines for reference.}}  
\label{figures4}
\end{figure}

\begin{figure}
\ \vspace{.4in}\\ 
\includegraphics[width=3.in]{Figures/fit.eps}
\caption{\rc{Plots of $F_{s,I}(t)= F_I(t)/\hbar$ (in the unit of ${\rm ps^{-1}}$) versus time  for the two different cases B and D of Table 3 in the main text for azulene, shown as black lines.  The data without bath are provided as red dashed lines for reference.}}  
\label{figures5}
\end{figure}

\rc{We confirmed numerically that the approximations, Eqs. (\ref{eq:kr-delta}) and (\ref{eq:ki-delta}), converge for sufficiently small $\delta \omega$ and large $M$.   For the calculations provided in the main text, we chose $\delta \tilde \nu=\delta \omega/(2\pi c)=0.1 \ {\rm cm^{-1}}$ and $M=50,000$. }
 \rc{Comparison of ${\mathcal K}_R(t)$ shown in Fig. 6 of main text and ${\mathcal K}_I(t)$ shown in  Fig. \ref{figures1} confirms that these results all approach the values without the effect of bath in the short time limit.  }

\rc{Similar approaches were used to decompose ${\mathcal J}_D(\omega)$ and ${\mathcal J}_F(\omega)$ into Ohmic parts and to calculate the contributions of the remainder as described above. Figures (\ref{figures2})-(\ref{figures5}) show real and imaginary parts of the resulting $D(t)$ and $F(t)$ in the short time limit for the two cases B and D of Table 3 in the main text. These results all confirm that these correlation functions decay fast as the strength of bath increases (except for slow logarithmic increase for $F_{I}(t)$) whereas they all approach correct bath-free values in the short time limit.  }

\ \vspace{5in}\\
\newpage
\newpage
%

\end{document}